\documentclass{aa}

\usepackage{graphicx,natbib}
\newcommand{\pun}[1]{\mbox{\rm\,#1}} 

\newcommand{\logg}{\ensuremath{\log g}}

\newcommand{\mlp}{\ensuremath{\alpha_{\mathrm{MLT}}}}

\newcommand{\Teff}{\ensuremath{T_{\mathrm{eff}}}}
\newcommand{\tauross}{\ensuremath{\tau_{\mathrm{ross}}}}

\newcommand{\draftflag}{false}

\newcommand{\beq}{\begin{equation}}
\newcommand{\eeq}{\end{equation}}
\newcommand{\pdx}[2]{\frac{\partial #1}{\partial #2}}

\newcommand{\eref}[1]{\mbox{(\ref{#1})}}

\newcommand{\cp}{\ensuremath{c_{\mathrm{p}}}}

\newcommand{\Hp}{\ensuremath{H_{\mathrm{P}}}}

\newcommand{\MHD}{{\sc RHD}}
\newcommand{\LHD}{{\sc LHD}}
\newcommand{\PHOENIX}{{\sc PHOENIX}}
\newcommand{\ATLASVI}{{\sc ATLAS6}}
\newcommand{\MHDb}{{\sc\bf RHD}}
\newcommand{\LHDb}{{\sc\bf LHD}}
\newcommand{\PHOENIXb}{{\sc\bf PHOENIX}}
\newcommand{\ATLASVIb}{{\sc\bf ATLAS6}}

\newcommand{\mosm}{A-3D}
\newcommand{\mobg}{B-3D}
\newcommand{\mosu}{S-3D}
\newcommand{\molo}{C-2D}
\newcommand{\mohi}{D-2D}

\newcommand{\fex}{\ensuremath{f_\mathrm{exchange}}}

\newcommand{\changed}{}

\begin{document}

\title{Numerical simulations of surface convection in a late M-dwarf}
\titlerunning{Numerical simulations of surface convection in a late M-dwarf}
\authorrunning{Ludwig et al.}
\offprints{Hans-G\"unter Ludwig}

\author{ Hans-G\"unter Ludwig\inst{1,2} \and France Allard\inst{2}
  \and Peter H. Hauschildt\inst{3,2}}
 
\institute{ Lund Observatory, Box~43, 22100 Lund, Sweden\\
            \email{hgl@astro.lu.se}
  \and 
  C.R.A.L., \'Ecole Normale Sup\'erieure, 69365 Lyon Cedex 7, France
  \and
  Dept. of Physics and Astronomy \&\ Center for Simulational Physics,
  University of Georgia, \\ Athens, GA 30602-2451
}

\date{Received date; accepted date}

\abstract{Based on detailed 2D and 3D numerical
radiation-hydrodynamics (RHD) simulations of time-dependent
compressible convection, we have studied the dynamics and thermal
structure of the convective surface layers of a prototypical late-type
M-dwarf ($\Teff\approx 2800\pun{K}$, $\logg=5.0$, solar chemical
composition).  The RHD models predict stellar granulation
qualitatively similar to the familiar solar pattern. Quantitatively, the granular
cells show a convective turn-over time scale of $\approx 100\pun{s}$,
and a horizontal scale of $80\pun{km}$; the relative intensity
contrast of the granular pattern amounts to 1.1\pun{\%}, and
root-mean-square vertical velocities reach 240\pun{m/s} at
maximum. Deviations from radiative equilibrium in the higher, formally
convectively stable atmospheric layers are found to be insignificant
allowing a reliable modeling of the atmosphere with 1D standard model
atmospheres. A mixing-length parameter of \mbox{\mlp=2.1} provides the
best representation of the average thermal structure of the RHD model
atmosphere while alternative values are found when fitting the
asymptotic entropy encountered in deeper layers of the stellar
envelope \mbox{(\mlp=1.5)}, or when matching the vertical velocity
field \mbox{(\mlp=3.5)}. The close correspondence between RHD and
standard model atmospheres implies that presently existing
discrepancies between observed and predicted stellar colors in the
M-dwarf regime cannot be traced back to an inadequate treatment of
convection in the 1D standard models. The RHD models predict a modest
extension of the convectively mixed region beyond the formal
Schwarzschild stability boundary which provides hints for the
distribution of dust grains in cooler (brown dwarf) atmospheres.
} 

\maketitle

\section{Introduction}

Late-type M-dwarfs are fully convective stars where the convective
flows penetrate far into the atmospheres reaching optical depths as
low as $10^{-3}$ \citep{Allard+Hauschildt95}. \citet{Allard+al97} have
reviewed the physical, spectroscopic, and photometric properties of
these objects.  In the past, model atmospheres have typically failed
to reproduce their spectroscopic and photometric properties in two
respects: i) the near-IR spectral distribution ($JHK$ colors) where,
independent of the source of water vapor line data used, models all
agree to predict an underluminous $K$-band (relative to $J$), and ii)
the optical M$_V$ vs $V-I$ color-magnitude relation, where all models
systematically predict bluer colors (i.e. being overluminous in $V$)
than observed.

\citet{Brett95} raised  the possibility that this  near-IR problem was
due to models being ``too cool in the upper photospheric layers'', and
suggested  two  possible  causes:  chromospheric  heating  and/or  the
treatment of convection based on mixing-length theory (MLT).

Hydrodynamical simulations of solar and stellar granulation including
a realistic description of radiative transfer have become an
increasingly powerful and handy instrument for studying the influence
of convective flows on the the structure of late-type stellar
atmospheres as well as on the formation of their spectra
\citep[e.g.][]{Nordlund+Dravins90, Steffen+Freytag91, Ludwig+al94,
Freytag+al96, Stein+Nordlund98, Asplund+al00}. Hitherto, model
calculations have been exclusively performed for atmospheres where
atomic lines are dominating the line blanketing. A possible next step
in the development of hydrodynamical models is towards cooler
atmospheres where {\em molecular absorption\/} dominates the
atmospheric energy balance. Constructing hydrodynamical model
atmospheres for cooler stars can shed light on the presently pressing
shortcomings of the classical models mentioned above. Regarding the
considerable improvements in the quality of the molecular opacities
and related atmospheric models, it becomes more and more important to
determine whether the treatment of convection by MLT is at the origin
of the observed discrepancies.

The basic questions we want to answer in this theoretical
investigation are: Is mixing-length theory adequate to handle
convection in the atmospheres of M-dwarfs?  And if so, which
mixing-length parameter~\mlp\ is necessary to reproduce the various
thermal and dynamical properties of an atmosphere (temperature profile
in the line forming region, surface boundary condition connecting to
stellar evolution models, convective velocities)?  We start by
describing some methodological aspects and the applied computer codes,
in particular discuss the critical question of how accurately we can
describe the complex radiative transfer within the hydrodynamical
simulations. We continue by presenting our results which give some
insight in what granulation looks like on the surface of an
M-dwarf. We proceed with quantitative estimates of the mixing-length
parameter, and discuss the consequences for conventional atmosphere
modeling. Finally, we extrapolate slightly beyond the existing
hydrodynamical models proper, and suggest a scenario for the transport
of dust grains in brown dwarf atmospheres due to convective overshoot
which is motivated from our present simulations at hotter
temperatures. Often we refer to the Sun as our benchmark for
comparison and assume some familiarity with its atmospheric
properties.

\section{Methodological aspects, computer codes}

The aim is to model the atmospheric structure of a prototypical late
M-dwarfs as realistically as possible, with a focus on the interplay
of convective flows and radiative transfer. Being well aware of the
limitations in our models, we took, whenever possible, a
\textit{differential} approach in trying to reduce the influence of
systematic uncertainties on the outcome of our investigation. This
concerned mostly the dimensionality of the problem: the
multi-dimensional, time dependent approach adopted in the
hydrodynamical simulations versus the one-dimensional, time
independent approach adopted in classical stellar atmospheres. We
ensured that the numerical treatment (i.e. implemented microphysics,
representation of radiative energy transport) in the two ``worlds''
was as similar as possible.  We employed various computer codes whose
names and main characteristics we introduce below. We elaborate on
specific aspects critical for the investigation in more detail later.

\MHDb: A radiation-hydrodynamics code developed by {\AA}.~Nordlund and
R.F.~Stein \citep[see][ and references therein]{Stein+Nordlund98} for
modeling stellar atmospheres in two or three spatial dimensions. It
implements a consistent treatment of compressible gas flows together
with non-local radiative energy exchange. The radiative transfer is
treated in LTE approximation, the wavelength dependence of the
radiation field is represented by a small number of wavelength bins
(see below). Open lower and upper boundaries, as well as periodic
lateral boundaries are assumed. The effective temperature of a model
(i.e. the average emergent radiative flux) is controlled indirectly by
prescribing the entropy of inflowing material at the lower boundary.
Magnetic fields are neglected.

\LHDb: A 1D Lagrangian hydrodynamics code developed by one of the
authors (HGL) used to calculate standard stellar atmospheres which can
be compared with results obtained with \MHD. Besides the reduced
dimensionality, the adopted physics (equation of state, radiative
transfer scheme) is the same as in \MHD. The convective energy
transport is based on MLT. In this paper we adopt the MLT formulation
of \citet{Mihalas78}.  Excluding one exception, all values of the
mixing-length parameter are given with reference to Mihalas'
formulation.

\PHOENIXb: A 1D model atmosphere code developed by two of the authors
\citep[PHH \&\ FA, for a detailed description
see][]{Hauschildt+Barron99}.  It implements a treatment of the
wavelength dependence of the radiation field with high resolution
based on direct opacity sampling. In this investigation \PHOENIX\
served as opacity data base, and was used for assessing the
quality of the simplified radiative transfer employed in the
hydrodynamics codes.

\ATLASVIb: A version of the 1D model atmosphere code developed by
R.~Kurucz \citep{Kurucz79}. It served as additional opacity data base.

\begin{table*}
\begin{flushleft}
\caption[]{%
The hydrodynamical models discussed in the paper: \textbf{Name} is
used to identify a model, \textbf{Dim.} the dimensionality of the
model, \textbf{Mesh} the number of grid points (XxYxZ, where Z denotes
the vertical, X and Y the horizontal directions), \textbf{Size} the
geometrical size of the computational domain, \textbf{Opacities} the
employed opacity source, \textbf{$N_\mathrm{OBM}$} the number of OBM bins
(``Opacity Binning Method'', for a description see Sect.~\ref{s:obm})
for calculating the radiative transfer, \Teff\ the resulting effective
temperature of the model including an estimate of its RMS
fluctuations, \logg\ the gravitational acceleration, and
\textbf{Modelcode} an internal model identifier.}
\label{t:mhdmodels}
\begin{tabular}{lclllclll}
\hline\noalign{\smallskip}
Name & Dim. & Mesh & Size\,[km] & Opacities & $N_\mathrm{OBM}$ & \Teff\,[K] & \logg & Modelcode \\
\noalign{\smallskip}
\hline\noalign{\smallskip}
\mosm & 3 & 125x125x82 & 250x250x87    & \PHOENIX & 4 & $2789.1 \pm 0.7$ & 5.0 & d3gt30g50n18 \\
\mobg & 3 & 250x250x82 & 500x500x116   & \PHOENIX & 4 & $2799.8 \pm 0.2$ & 5.0 & d3gt30g50n19 \\
\mosu & 3 & 125x125x82 & 6.0x6.0x3.2 (Mm)  & Uppsala & 4 & $5640 \pm 14$ & 4.44 & sun3d \\
\molo & 2 & 125x82     & 250x80.3  & \ATLASVI & 1 & $2774 \pm 1.6$ & 5.0 & d2gt30g50n8 \\ 
\mohi & 2 & 251x162    & 250x80.3  & \ATLASVI & 1 & $2770 \pm 1.6$ & 5.0 & d2gt30g50n9 \\
\noalign{\smallskip}
\hline
\end{tabular}%
\end{flushleft}
\end{table*}

Table~\ref{t:mhdmodels} summarizes the properties of the
hydrodynamical models discussed in the paper. Model~\mosm\ is our
M-dwarf reference model. The twice as large model~\mobg\ was primarily
calculated for checking effects of the domain size, the solar model
~\mosu\ was added for assessing scaling properties with effective
temperature and gravitational acceleration. We note that the solar
model is not strictly differentially comparable to the M-dwarf models
since it is employing different opacity sources and equation of state
which stem from the Uppsala stellar atmosphere package
\citep{Gustafsson+al75}. We do not consider this as particularly
critical since the physical conditions in the atmospheres of M-dwarfs
and the Sun are so different that one looses the advantages of a
differential approach anyway. The 2D models~\molo\ and~\mohi\ are
models which were considered in the forefield to investigate effects
of the numerical resolution. They are based on \ATLASVI\ opacities
without contributions of molecular lines and employ grey radiative
transfer. Due to the different input physics their behavior is
qualitatively different from the more realistic 3D models. Despite
their shortcomings, they are of interest for qualitatively
understanding the interplay of convection and radiative transfer in
optically thin regions, and therefore will be discussed in more
detail.  As is apparent from the table, all hydrodynamical M-dwarf
models show very small fluctuations around their average effective
temperature. This reflects the fact that the horizontal and temporal
fluctuations of all quantities are small compared to the Sun --- a
central feature of the atmospheres of late-type M-dwarfs.

\subsection{The equation of state}

\begin{figure}
\resizebox{\hsize}{!}{\includegraphics[draft = \draftflag]%
{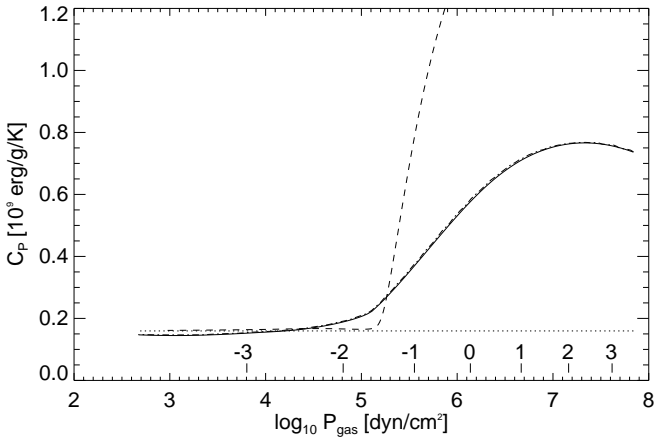}}
\caption[]{%
Comparison of the specific heat at constant pressure from various
equations of state in a representative M-dwarf: \MHD\
(\textbf{solid}), and \PHOENIX\ (\textbf{dash-dotted}). If one
artificially removes the contribution of $\mathrm{H}_2$ molecular
formation \cp\ is largely reduced (\textbf{dotted}). For further
comparison the run of \cp\ in a solar model atmosphere is depicted
(\textbf{dashed}).  For the M-dwarf model a logarithmic Rosseland
optical depth scale is indicated by the tick marks near the abscissa.
\label{f:cps}
} 
\end{figure}

Figure~\ref{f:cps} shows that the equation of state (EOS) which is
employed in the hydrodynamical simulations and \LHD\ models is very
similar to the \PHOENIX\ EOS. We do not expect significant systematic
differences by applying these two different equations of state in our
various model calculations.  The equations of state of \MHD\ and
\PHOENIX\ treat ionization and molecular formation assuming
Saha-Boltzmann statistics. Since non-ideal effects are not pronounced
in atmospheres of M-dwarfs the inclusion of $\mathrm{H}_2$ molecular
formation is the main ingredient required to obtain a realistic
description of the thermodynamics of the stellar plasma.

Indeed, Fig.~\ref{f:cps} demonstrates that in the M-dwarf atmosphere
$\mathrm{H}_2$ molecular formation is the most important contributor
for increasing the specific heat above the value associated with the
purely translatorial degrees of freedom.  In the Sun the dominant
contributor is the hydrogen ionization which has an even more dramatic
effect on the heat capacity of the stellar plasma. I.e. from the
perspective of the content of latent heat in the gas flows M-dwarf
atmospheres are not particularly extreme objects.

\subsection{Radiative transfer}
\label{s:obm}

\begin{figure}
\resizebox{\hsize}{!}{\includegraphics[draft = \draftflag]%
{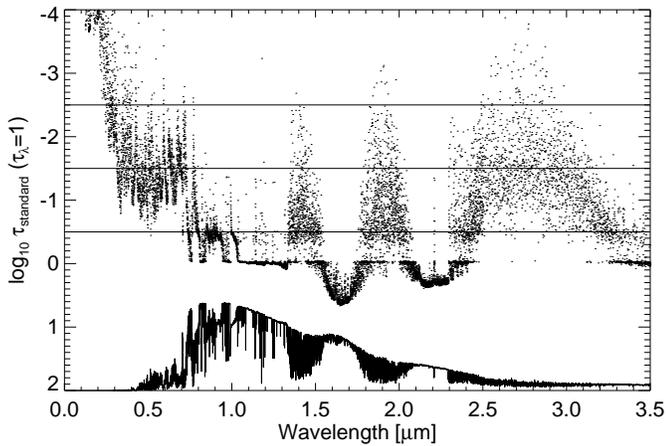}}
\caption[]{%
Scatter plot of standard optical depth where the monochromatic optical
depth reaches unity as a function of wavelength in a \PHOENIX\ model
atmosphere with \Teff=2900\pun{K}, \logg=5.3, and solar chemical
composition. Standard optical depth is the optical depth scale in the
continuum plus opacity contributions of $\mathrm{H}_2$ and
$\mathrm{H}_2\mathrm{O}$ molecular lines at
$\lambda=1.2\pun{$\mu$m}$. The horizontal lines mark the thresholds
which were used for defining the OBM opacity bins (see text). The
curve in the lower part of the panel is the emergent flux (arbitrarily
scaled) of the model.
\label{f:bands}
} 
\end{figure}

An important problem when modeling M-dwarf envelopes is the
treatment of the large number of absorption lines in their atmospheres
which are mostly of molecular origin. The complex wavelength
dependence of the radiation field is illustrated in
Fig.~\ref{f:bands}. While it is already a formidable task to treat the
frequency dependence of the radiation field in 1D model atmospheres,
this is even more the case in hydrodynamical models where one has to
account for the 3D geometry of the flow field and its temporal
evolution. Present computer capacity allows only for a very modest
number of frequency points to be included the modeling of the
radiation field within a hydrodynamical simulation. However, the
situation is somewhat alleviated. For the interaction of
hydrodynamics and radiative transfer only the frequency integrated
net amount of radiative heating (or cooling if negative)
\beq
Q_\mathrm{rad} = 4\pi\rho \int_0^\infty \chi_\nu(J_\nu-S_\nu)\, d\nu
\eeq 
($\rho$ denotes the density, $\chi_\nu$ the monochromatic absorption
coefficient, $J_\nu$ the angular average of the intensity, $S_\nu$ the
source function) is of relevance. I.e. we do not need to know the
detailed frequency dependence of the radiative heating. Perhaps
surprisingly, one can obtain a reasonably accurate description of the
radiative heating with very few frequency points by a judicious choice
of the frequency discretization: frequencies with a similar run of
monochromatic optical depth are grouped together into opacity
``bins''. In the context of hydrodynamical stellar atmospheres this
{\em Opacity Binning Method (OBM)\/} was originally introduced by
\citet{Nordlund82}, and some variants and refinements were
subsequently developed \citep{Ludwig+al94}. The basic idea is to group
frequencies which reach monochromatic optical depth unity within a
certain depth range of the atmosphere into one frequency
bin. Effectively, this combines frequencies into the same bin which
belong to the continuum, weaker, and stronger spectral lines. The
opacity in each bin is an average (Rosseland or Planck) over the
frequency range which is represented by that bin. With these average
opacities, supplemented by the frequency integrated source function,
the equation of radiative transfer has to be solved only once per bin
to obtain the formal solution of the radiative transfer for a given
flow field. The binning procedure is illustrated in Fig.~\ref{f:bands}
where the horizontal lines separate the depth ranges --- as measured
by a frequency independent standard optical depth --- which are sorted
into the various OBM bins. In the present investigation we used 4 bins
in our non-grey hydrodynamical models, representing the continuum and
line wings, as well as the cores of weak, medium strong, and strong
spectral lines.

It is clear that the sorting procedure described before is specific to
the stellar atmosphere under consideration and has to be repeated as
soon as the atmospheric parameters differ widely from the one where
the sorting was done. We used a model atmosphere calculated with
\PHOENIX\ at $\Teff=2900\pun{K}$ and $\logg=5.3$ as reference
atmosphere for the grouping. It is sufficiently close to the
atmospheric parameters of the hydrodynamical models. 
{\changed
This has been checked by studying the performance of the sorting when
applied to differing atmospheric parameters (\logg\ down to
3.0, and \Teff\ up to 3300\pun{K}).}

\begin{figure}
\resizebox{\hsize}{!}{\includegraphics[draft = \draftflag]%
{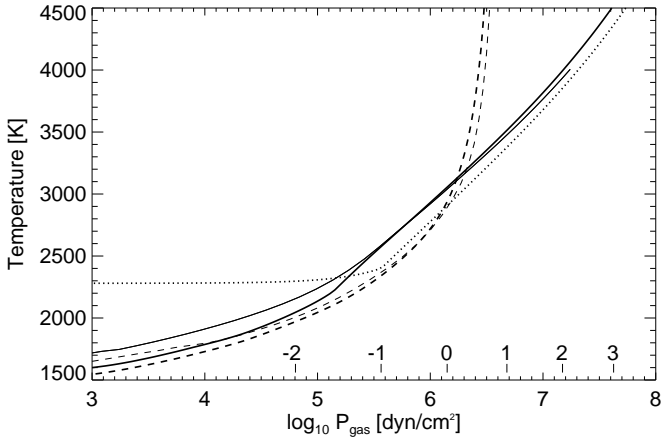}}
\caption[]{%
Comparison of 1D standard model atmospheres (\Teff=2800\pun{K},
\logg=5.0) in purely radiative (\textbf{dashed} lines) and
radiative-convective ($\mlp=1.0$, \textbf{solid} and \textbf{dotted}
lines) equilibrium. The \textbf{thick} lines are based on \LHD\
atmospheres using the approximate OBM radiative transfer, the
\textbf{thin} lines are based on \PHOENIX\ models using opacity
sampling with high wavelength resolution. The \textbf{dotted} line
depicts an \LHD\ model employing grey radiative transfer. A
logarithmic Rosseland optical depth scale is indicated by the tick
marks near the abscissa.
\label{f:trt}
} 
\end{figure}

The sorting criterion that $\tau_\lambda=1$ should fall within a
certain depth range in the atmosphere does not guarantee that the
overall depth dependence of $\tau_\lambda$ is similar for all
frequencies grouped together --- a condition for allowing the
interchange of the solution of the transfer equation with the
frequency integration when evaluating $Q_\mathrm{rad}$. In particular,
the simultaneous presence of atomic and molecular lines can lead to
significantly different functional forms of the monochromatic optical
depth at different frequencies: deeper regions of the atmosphere too
hot to allow for molecule formation might be dominated by atomic lines
while higher and cooler regions which allow for molecule formation
might be dominated by molecular lines. If the atomic and molecular
lines emerge from different elements there is no physical
connection between them leading to an uncorrelated behavior in optical
depth of deeper and higher layers with frequency. Such a situation
would be unfavorable for the OBM. {\changed However, the OBM is rather
successful in reproducing the heat exchange between radiation and
matter --- as evident from Fig.~\ref{f:trt}}. 
This is linked to the statistical dominance of molecular
absorption in {\em all\/} radiative layers of the rather cool
atmosphere under consideration. 

{\changed Another point concerning the present implementation} of the
OBM is our usage of global Rosseland means --- i.e. Rosseland averages
over the whole frequency range --- for representing the average
opacity in each bin. For the the continuum we took the Rosseland means
themselves while we scaled them by factors of $10^1$, $10^2$ , and
$10^3$ for the bins representing the successively stronger lines. The
scaling factors correspond to the factors of 10 in optical depth
selected as thresholds for the sorting procedure which are
$\log_{10}\tau_\mathrm{standard}=-0.5, -1.5,\,\,\mathrm{and}\, -2.5$
(see Fig.~\ref{f:bands}). The basic assumption behind this approach is
that the line opacity scales with temperature and pressure like the
continuous opacity. The source function has been integrated over the
frequency ranges of the individual bins. While it is certainly not the
optimal representation of the opacities it was dictated by the lack of
a detailed tabulation of the monochromatic opacities over the full
temperature-pressure range of interest. Work is presently under way to
generate such tabulation which is a non-trivial task due to the
enormous amount of line data which need to be processed.
{\changed Considering the various approximations described before one
might ask whether the OBM is a real improvement beyond a grey
description. Eventually, this can be tested quantitatively by
comparing 1D atmospheres computed with high frequency resolution or
the OBM. Such a comparison serves as ultimate indicator of the
performance of the OBM.}

Figure~\ref{f:trt} shows a comparison of 1D model atmospheres
calculated with \PHOENIX\ employing opacity sampling with a wavelength
resolution of $\approx 2\pun{{\AA}}$, and \LHD\ employing the OBM as
described before. Comparing flux constant models in radiative and
radiative-convective equilibrium shows a good correspondence of the
resulting equilibrium temperature profiles. While the radiative
equilibrium models are less important for the investigation of M-dwarf
atmospheres, they were added to the comparison to show the similarity
in the radiative transport properties independent of influences of the
convective transport. More important are the models in
radiative-convective equilibrium since they are closer to the actual
physical situation. For judging the correspondence between opacity
sampling and OBM models one should take the grey model as benchmark
which represents a strongly simplified (equivalent to one frequency
bin) description of radiative transfer. Clearly, the OBM matches much
more closely the \PHOENIX\ opacity sampling model. The cooling of the
higher atmosphere by lines is reasonably represented as well as the
backwarming of the deeper layers.

The most important deviation between opacity sampling and OBM model
occurs in the layers where the transition from convectively to
radiatively dominated energy transport takes place (around $\log P
=5.3$). The OBM stratification becomes noticeably cooler. This was
traced back to an insufficient heating of the gas in the OBM continuum
bin. This in turn is probably related to the use of Rosseland averages
for the continuum opacity in these optically thin regions. Planck
averages would be more suitable.  But again, at the time this work was
performed only globally averaged opacities were available. Rosseland
and Planck averages differ by at least a factor of 100 in these
regions, making an ad-hoc switching from one to the other
problematic. However, we consider the remaining deviations not as
vital, in particular with respect to the calibration of the
mixing-length parameter which we describe later in this paper.  The
calibration is a result of a differential comparison of models which
{\em all\/} base on the OBM.

For completeness, we finally remark that only the exchange of energy
was considered within the radiative transfer. The exchange of momentum
was neglected; the prevailing relatively high mass densities combined
with low radiative fluxes render radiation pressure unimportant for
the structure of M-dwarf atmospheres.

\section{Characteristic time scales}

\begin{figure}
\resizebox{\hsize}{!}{\includegraphics[draft = \draftflag]%
{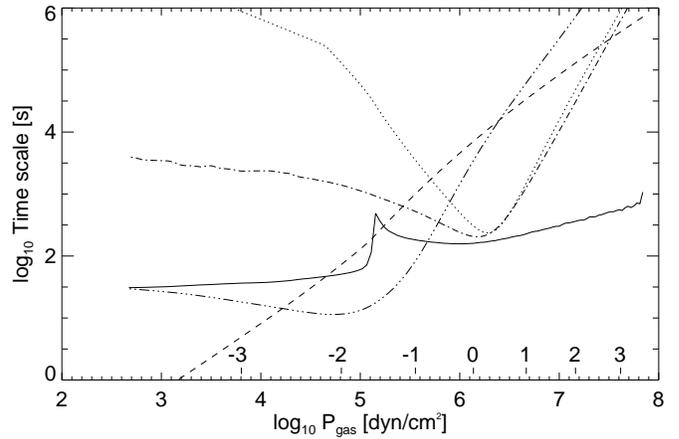}} 
\caption[]{%
Various characteristic time scales as a function of pressure in an
M-dwarf model: Brunt-V\"ais\"al\"a period (\textbf{solid}),
Kelvin-Helmholtz time scale (\textbf{dashed}), radiative relaxation
time based on \ATLASVI\ Rosseland opacities (\textbf{dotted}),
\PHOENIX\ Rosseland (\textbf{dashed-dotted}), and \PHOENIX\ Planck
(\textbf{triple-dot-dashed}) opacities. A logarithmic Rosseland
optical depth scale is indicated by the tick marks near the
abscissa. The kink in the run of the Brunt-V\"ais\"al\"a period is
located at the boundary between the convectively stable and unstable
part of the model.
\label{f:ctimes}
} 
\end{figure}

To get an overview of the problem Fig.~\ref{f:ctimes} shows various
characteristic time scales in a representative M-dwarf model (\LHD\
model with \Teff=2790\pun{K}, \logg=5.0, \mlp=1.0). The expressions
which we applied in the calculation of the time scales are summarized
in the appendix. The time scales have been evaluated under simplifying
assumptions, and should therefore be taken as order of magnitude
estimates only.

The radiative time scales based on \ATLASVI\ and \PHOENIX\ Rosseland
mean opacities are rather similar in the deeper layers while being
very different in the optically thin regime. This emphasizes the great
influence of molecular absorption which becomes important at cooler
temperatures and which is not included in the \ATLASVI\ opacities. The
radiative time scales from \PHOENIX\ Rosseland and Planck opacities
differ also significantly. Comparing the radiative time scales to a
dynamical or convective time scale as given by the
Brunt-V\"ais\"al\"a period leads us to expect that one gets a
qualitatively different behavior depending on the treatment of the
radiative transfer. We will see that models based on a
frequency-independent Rosseland opacities show significant deviations
from radiative equilibrium conditions. A more realistic treatment ---
in the optically thin regions more closely represented by the Planck
mean opacities --- results in an atmospheric structure closer to
radiative equilibrium. In all cases, one expects an almost adiabatic
structure in the deeper layers since the radiative relaxation time is
much longer than the convective time scale.

\subsection{A remark on thermal relaxation}

\MHD\ uses an explicit numerical scheme for advancing the solution in
time. It is well known that in an explicit scheme the time step is
limited to a fraction of the dynamical time scale, more precisely to
the limit given by the Courant-Friedrichs-Levy stability
condition. Together with the amount of available computer --- and
wallclock --- time, this limits the time interval which can be covered
by a 3D \MHD\ model to about $10^4\pun{s}$ of stellar time.
Figure~\ref{f:ctimes} seemingly implies that it is impossible to
obtain a thermally relaxed hydrodynamical model within this time
interval since the Kelvin-Helmholtz time scale of the deeper layers is
at least an order of magnitude larger. Contrary to the impression
given in Fig.~\ref{f:ctimes}, in the multi-dimensional hydrodynamical
simulations this does not pose a problem since the thermal relaxation
of the model is {\em not\/} governed by the time scale for the
exchange of energy as expressed by the Kelvin-Helmholtz time. In fact,
the thermal evolution of the deeper layers of the \MHD\ models is
governed by the time scale for the exchange of mass in these layers
which is much shorter (as quantified below, see
Fig.~\ref{f:massex}). Due to the exponential run in density of the
atmosphere the mass exchange consists primarily of the replacement of
mass by fresh material stemming from deeper layers. In the
hydrodynamical model it is ultimately fed into the computational
domain at the lower boundary.  
{\changed The convective energy flux is the net effect of the energy
transported by counteracting, opposing mass currents in and out of a
test volume.  The advective --- as opposed to diffusive --- nature of
the mass transport leads to a situation where there can be a large
imbalance in the energy content of the mass currents: the in-coming
mass elements carry the energy (strictly speaking the specific
entropy) of much deeper layers (in the hydrodynamical model the
entropy at the lower boundary), while the out-going elements just
carry the local energy density.  This implies an energy flux much
larger than the one close to equilibrium conditions. The system is
driven much faster towards equilibrium than implied by the
classical Kelvin-Helmholtz time scale, which assumes an energy flux
as encountered close to equilibrium conditions.}

We have argued from the perspective of our hydrodynamical models. But
the fast thermal relaxation is driven by a physical mechanism implying
that a real stellar convection zone does not operate any
differently. Hence, we want to stress that to our understanding the
thermal relaxation of a convective layer is generally {\em not\/}
governed by the classical Kelvin-Helmholtz time scale but by the
usually shorter time scale of mass exchange.

In the one-dimensional \LHD\ models no mass exchange takes place and
the convective energy flux is computed from MLT, {\changed in which
convection is modeled as diffusion process}. Under such
circumstances the time scale of the thermal evolution of the
convective layers is indeed comparable to the Kelvin-Helmholtz time
scale as shown in Fig.~\ref{f:ctimes}. Of course, covering the time
interval necessary for the slower thermal relaxation in a 1D model run
poses no problems due to the largely reduced computational costs.

\section{2D precursor models}

\begin{figure}
\resizebox{\hsize}{!}{\includegraphics[draft = \draftflag]%
{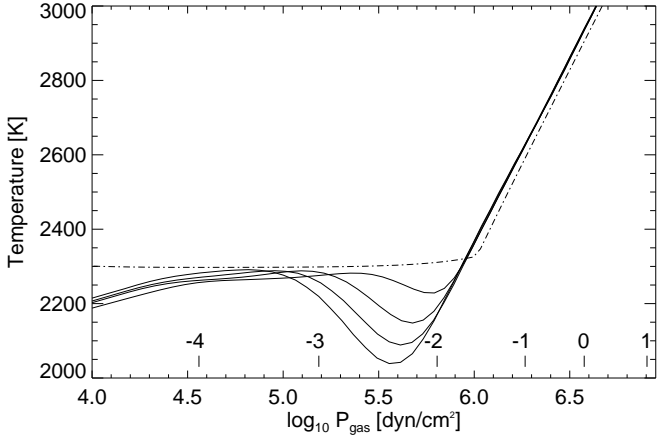}}
\caption[]{%
Temporal evolution of the average temperature profile of model~\molo\
in steps of 5\pun{ks} showing a successive cooling of the layers
around $\log P = 5.5$ (\textbf{solid} lines). For comparison a \LHD\
model with $\mlp=1.0$ is plotted (\textbf{dash-dotted}). All models
are based on grey radiative transfer employing \ATLASVI\ opacities
devoid of molecular contributions. All models have the same
atmospheric parameters (\Teff=2770\pun{K}, \logg=5.0). A logarithmic
Rosseland optical depth scale is indicated by the tick marks near the
abscissa.
\label{f:evo2d}
} 
\end{figure}

Models~\molo\ and~\mohi\ are 2D models (see Tab.~\ref{t:mhdmodels})
which were constructed to gain experience regarding the size of the
computational domain and grid resolution. They are based on grey
radiative transfer utilizing \ATLASVI\ opacities which do not include
contributions of molecular lines.  The choice of the \ATLASVI\
opacities was dictated by the lack of more realistic opacities during
the initial phase of the project.  From Fig.~\ref{f:ctimes} one might
readily conclude that their atmospheric structure will be dominated by
convection since the radiative relaxation times in the atmosphere are
long as compared to the dynamical time scale. Indeed, when starting
from a temperature structure taken from a \LHD\ model based on MLT, we
find a rapid cooling of the originally radiatively stratified
atmospheric layers by convective overshooting (see
Fig.~\ref{f:evo2d}). Convection tends to transform the stratification
into a purely adiabatic one since the radiative heating is too weak to keep
the temperature close to the radiative equilibrium value. The
transformation into an almost adiabatic stratification takes too long
to be covered within the multi-dimensional hydrodynamical simulations.
However, we conducted numerical experiments with \LHD\ where an
\textit{ad-hoc} velocity field mimicking the convective overshoot was
put into the atmospheric regions which are formally stable according
to the Schwarzschild criterion. The \LHD\ models indicate that the
asymptotic temperature profile tends to a fully adiabatic
stratification.

Due to the unrealistic opacities, the models cannot give a good
representation of a real M-dwarf atmosphere. They are nevertheless
interesting from a numerical point of view. The resolution of the less
resolved model~\molo\ is already sufficient to represent the
convective transport properties, in particular in the important
transition region from convectively to radiatively dominated energy
transport: after an initial relaxation phase we find in both models
that the minimum temperature drops linearly with a rate of
$\frac{dT}{dt}=-0.011\pun{K/s}$.  
{\changed Looking at further diagnostics at comparable instances
during the secular evolution of the model runs we observe that RMS
velocities are similar within a 20\pun{\%} level. The higher resolved
model shows somewhat more small-scale features, and its downflows
appear more concentrated. However, we are primarily interested in the
balance of convective to radiative energy transport. The temperature
drop rate is a convenient measure of the relative efficiency of both
processes.  Hence, we conclude from the similarity of the drop rates
that the resolution of the less resolved model applied in 3D
simulations is sufficient to model the transport properties of the
convective flows.}

Another conclusion which can be drawn from these models is that
standard MLT models can give quite misleading predictions of the
atmospheric temperature structure if the radiative relaxation time is
long in comparison to convective time scales. In other words, one
should be cautious when the coupling of the temperature structure to
the radiative equilibrium temperature is so weak that overshoot of low
amplitude happens essentially adiabatically. In the following we shall
see that M-dwarf atmospheres are unlikely to exhibit such conditions.

\section{3D models: results and discussion}

\subsection{General flow structure}

\begin{figure}
\resizebox{\hsize}{!}{\includegraphics[draft = \draftflag]%
{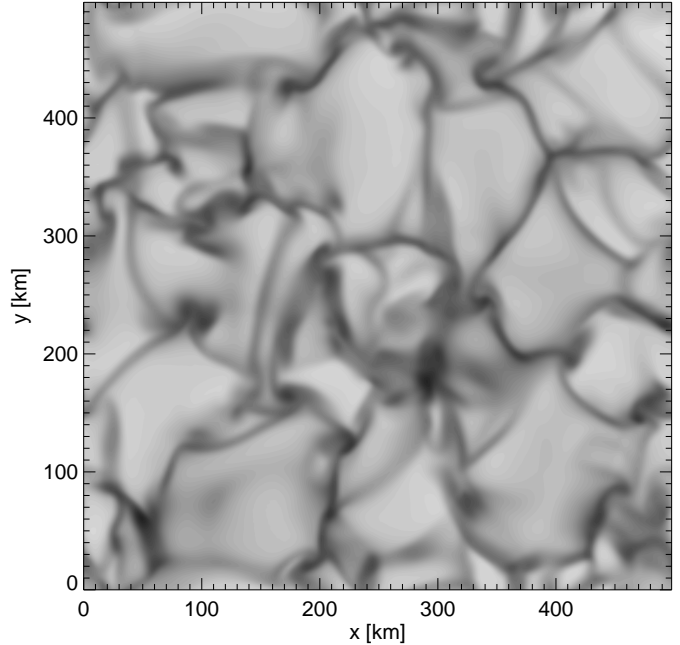}}
\caption[]{%
Typical snapshot of emergent intensity during the evolution of
model~\mobg. The intensity contrast amounts to $\delta
I_\mathrm{rms}=1.2\pun{\%}$ at this instant in time. Note, that the
contrast of the intensity pattern in the graphics is enhanced with
respect to its real appearance.
\label{f:flow}
}
\end{figure}

\begin{figure}
\resizebox{\hsize}{!}{\includegraphics[draft = \draftflag]%
{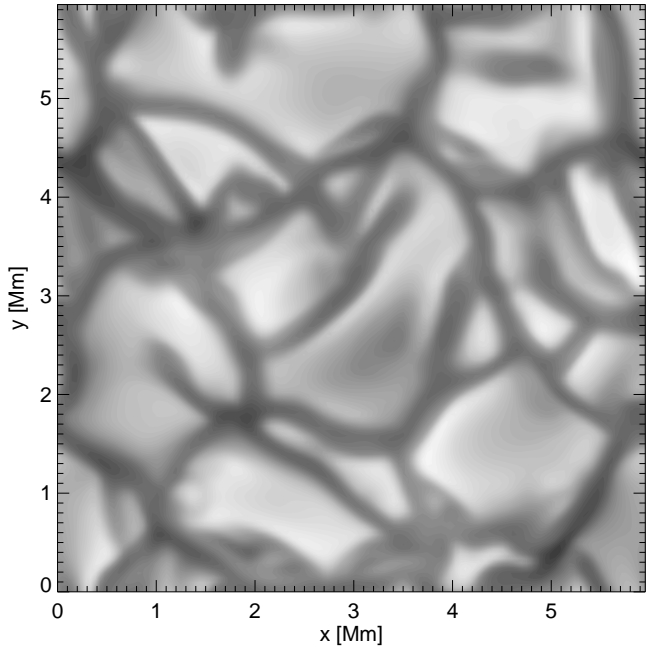}}
{\changed
\caption[]{%
Like Fig.~\ref{f:flow}, but for the {\em solar\/} model~\mosu. The
intensity contrast amounts to $\delta I_\mathrm{rms}=16.7\pun{\%}$ at
this instant in time. Note, that for reproduction purposes spatial as
well as intensity scale are altered in comparison to Fig.~\ref{f:flow}.
\label{f:flowsun}
} 
}
\end{figure}

{\changed %
Figure~\ref{f:flow} shows a typical snapshot of the emergent intensity
during the temporal evolution of model~\mobg. For comparison,
Fig.~\ref{f:flowsun} shows a similar snapshot from the solar
run~\mosu. Note, that the spatial and intensity scaling is very
different in the reproductions, hence, only relative geometrical
properties should be directly compared. The average relative RMS
intensity contrast of the granular pattern amounts to 1.1\pun{\%} in
the M-dwarf as opposed to 16\pun{\%} in the solar case.  The first
thing to note is that surface convection in an M-dwarf produces a
granular pattern qualitatively resembling solar granulation: bright
extended regions of upwelling material which are surrounded by dark
concentrated lanes of downflowing material. The dark lanes form an
interconnected network.  Looking more closely, granules are less
regularly delineated in M-dwarfs, the inter-granular lanes show a
higher degree of variability in terms of their strength.  A feature
which is uncommon in the solar granulation pattern are the dark
``knots'' (e.g. at $x=100\pun{km}$ and $y=230\pun{km}$ in
Fig.~\ref{f:flow}) found in or attached to the inter-granular
lanes. The knots are associated with strong downdrafts which carry a
significant vertical component of angular momentum.  The width of the
inter-granular lanes to the typical granular size is smaller in
M-dwarfs. Inspecting the velocity field (not shown) in vicinity of the
continuum forming layers shows less pronounced size differences. This
indicates that the relatively broader lanes in the solar case are the
result of a stronger smoothing of the temperature field due to a more
intense radiative energy exchange, i.e. the effective Pecl{\'e}t
number of the flow is larger around optical depth unity in M-dwarfs.}

\begin{figure}
\resizebox{\hsize}{!}{\includegraphics[draft = \draftflag]%
{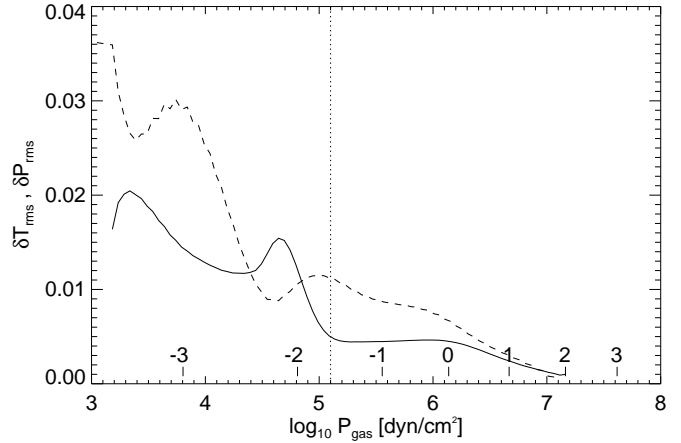}}
\caption[]{%
Relative horizontal temperature (\textbf{solid}) and pressure
(\textbf{dashed}) fluctuations of model~\mosm. The fluctuations in the
region $\log P < 5$ are likely to be overestimated in the model,
i.e. should be considered as upper limits. The \textbf{dotted} line indicates
the approximate location of the Schwarzschild boundary. A logarithmic
Rosseland optical depth scale is indicated by the tick marks near the
abscissa.
\label{f:flucts}
} 
\end{figure}

The different magnitude of the intensity contrast already indicates
that horizontal fluctuations of the thermodynamic quantities are small
in M-dwarf atmospheres. Figure~\ref{f:flucts} shows the run of the
relative temperature and pressure fluctuations in
model~\mosm. Plotted are long term temporal and horizontal
\footnote{The horizontal averaging is performed on the geometrical
depth scale. Due to the small horizontal fluctuations, alternative
choices (e.g. on the optical depth scale) give similar results.} RMS
averages. As we will argue later, the fluctuations in the higher
atmosphere are likely to be overestimated in the model. But even taken
at face value they are quite modest. The low level of fluctuations in
the thermodynamic quantities is accompanied by small flow velocities,
the maximum Mach number amounts to 6.5\pun{\%} in model~\mosm. MLT
models show that Mach numbers drop to even lower values as one goes to
lower effective temperatures which are encountered in the regime of
brown dwarfs. Our findings may have a direct bearing on the dust
formation conditions in such objects. Our hydrodynamical simulations
support the view that dust forming layers in cool main sequence
objects experience only small variations around their mean
thermodynamic state. This view is clearly at odds with the scenario
discussed by \citet{Helling+al01} who study the dust formation in
turbulent brown dwarf atmospheres. Helling et al. assume thermodynamic
fluctuations of order unity.

{\changed %
Another important feature, which distinguishes M-dwarf atmospheric
conditions from solar ones, is the extent of the convective layers. As
evident from Figures~\ref{f:talpha1} and~\ref{f:talpha2}, the
convective motions reach much lower optical depth
($\log\tau_\mathrm{Rosseland}\approx -1.5$) in the M-dwarf than in the
Sun ($\log\tau_\mathrm{Rosseland}\approx 0$). Two factors contribute
to this behavior. The temperature gradient in non-grey radiative
equilibrium is significantly steeper in the M-dwarf than in the Sun,
presumably due to efficient cooling of the atmospheric layers by
molecular lines. Furthermore, the $\mathrm{H}_2$ molecule formation
reduces the adiabatic gradient in the M-dwarf atmosphere, while in the
Sun the hydrogen recombination is essentially completed in
subphotospheric layers. The different radiative and thermodynamic
conditions favor the presence of convection in M-dwarf atmospheres.}

In the deeper layers we observe the tendency --- also known from solar
simulations \citep[see][]{Stein+Nordlund98} --- that the granular
network of downflows decays into isolated downdrafts. Our models are
rather shallow reaching only 2.3 pressure scale heights below optical
depth unity. Thus we cannot follow the change of flow topology as far
as has been done for solar models, but within the limited depth range
comprised by our models we do not see indications of a qualitatively
different behavior as found in the Sun.

\subsection{Horizontal scales}

\begin{figure}
\resizebox{\hsize}{!}{\includegraphics[draft = \draftflag]%
{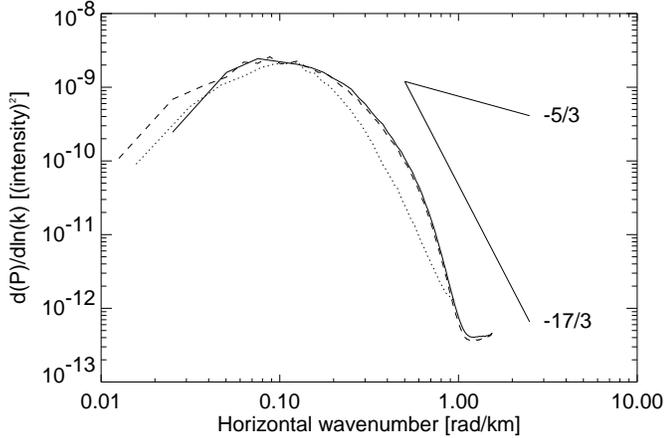}}
\caption[]{%
Power spectrum of the emergent intensity pattern for model~\mosm\
(\textbf{solid}), and \mobg\ (\textbf{dashed}), as well as the solar
model~\mosu\ (\textbf{dotted}).  The spectrum of the solar model was
scaled in power as well as wavenumber to match the peak of power in
the M-dwarf models. Lines with slopes $\frac{dP}{dk}\propto k^{-5/3}$
and $\frac{dP}{dk}\propto k^{-17/3}$ are shown for comparison with
expectations from simplistic turbulence theory. The total relative
intensity contrasts amount to $\delta I_\mathrm{rms}=$ 1.12\pun{\%}
(\mosm), 1.12\pun{\%} (\mobg), and 16.2\pun{\%} (\mosu).
\label{f:poweri}
} 
\end{figure}

\begin{figure}
\resizebox{\hsize}{!}{\includegraphics[draft = \draftflag]%
{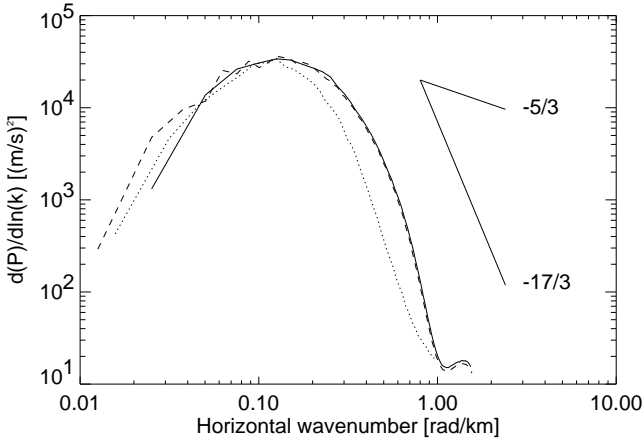}}
\caption[]{%
Power spectrum of the vertical velocity pattern, \textbf{line styles as in
Fig.~\ref{f:poweri}}. The solar power spectrum was scaled to match the
peak of power of the M-dwarf models.  The velocity pattern was
taken from the layer where the RMS vertical velocity reaches its
maximum, with absolute values of $v_\mathrm{rms,max}=$ 236\pun{m/s}
(\mosm), 241\pun{m/s} (\mobg), 2600\pun{m/s} (\mosu).
\label{f:powerv}
} 
\end{figure}

In the following we shall discuss spatial power spectra of intensity
and vertical velocity. Note, that in the Figures~\ref{f:poweri}
and~\ref{f:powerv} we display power per logarithmic wavenumber
interval and not per unit wavenumber --- the more common choice. This
allows visually for a more direct identification of the power carrying
scales. However (in humble respect of Kolmogorov's achievements), we
labeled the power laws drawn for comparison with the familiar spectral
index of power per unit wavenumber. The spectra are temporal averages
over one to a few convective turn-over times and many convective
cells, so that they are statistically representative.


Figure~\ref{f:poweri} shows a comparison of power spectra of the
emergent intensity (more precisely: the intensity in the OBM continuum
bin in vertical direction at the upper boundary of the computational
volume) for the models~\mosm, and \mobg\ as well as the solar
model. Figure~\ref{f:powerv} shows a corresponding comparison of the
vertical velocity component measured at the layer where its RMS value
reaches the maximum. To facilitate a comparison between the M-dwarf
models and the solar model, the solar model was arbitrarily scaled in
power and wavenumber so that the maxima and position of the power
distributions matched.
{\changed For the M-dwarf models the hump in power at the highest
wavenumbers is likely an artifact of an imperfect choice of parameters
controlling the small scale dissipation and should be ignored.}

Fluctuations in the Sun are an order of magnitude larger and had to be
reduced accordingly in order to match the height of the maxima in the
M-dwarf models, while the horizontal scale of the solar model had to
be shrunk by a factor of 15. This factor closely corresponds to
variation of the pressure scale height at the surface (factor 12.5)
and follows the scaling found for hotter models by
\citet{Freytag+al97}. Different from Freytag et al., we accounted for
the change of the mean molecular weight of the stellar gas since the
H$_2$ molecular formation leads to a significant increase in M-dwarf
atmospheres. The peak power of the M-dwarf models is located at an
absolute scale of 80\pun{km} for intensity and 63\pun{km} for velocity
structures.

Comparing the width of the power distribution at 1\pun{dex} below peak
level, the M-dwarf model~\mobg\ displays a range of scales of
1.4\pun{dex}, the solar model of 1.1\pun{dex}. As is evident from the
figures, the solar scale distribution is slightly but noticeably
narrower. 
{\changed This might be traced back to the lower P{\'e}clet number of
the plasma encountered at the surface of an M-dwarfs as opposed to
solar conditions. The formation of small scale structures is more
strongly suppressed by the radiation field under solar conditions,
leading to an overall narrower power distribution.}
Model~\mosm\ apparently lacks
some larger scale power indicating that the computational box is
somewhat too small in this case. However, since the total fluctuations
are very similar in the models we conclude that the relevant
structures are captured in model~\mosm\ which serves as our standard
model.

Towards high wavenumbers the M-dwarf spectra do not show a clear power
law behavior. In any case, they cannot be fitted with a $-5/3$ slope,
and are at best roughly compatible with a $-17/3$ slope.  We would
like to emphasize that despite the very different physical parameters
of M-dwarf atmospheres this is qualitatively what was already found in
solar models, and is at odds with expectations for homogeneous and
isotropic turbulence. While smaller scales in the hydrodynamical
models are definitely influenced by the artificial viscosity, it is
still somewhat surprising that the asymptotic behavior in the Sun as
well as in the M-dwarf is so similar despite their quite different
characteristic flow speeds (i.e. Mach numbers).  While our present
models cannot make definite statements about the properties of small
scale turbulence, a ``non-Kolmogorov'' behavior cannot be ruled out
neither. See \citet{Nordlund+al97} for an in-depth discussion of this
issue in a solar context.

\subsection{Mean temperature stratification and corresponding effective 
            mixing-length parameter}

\begin{figure}
\resizebox{\hsize}{!}{\includegraphics[draft = \draftflag]%
{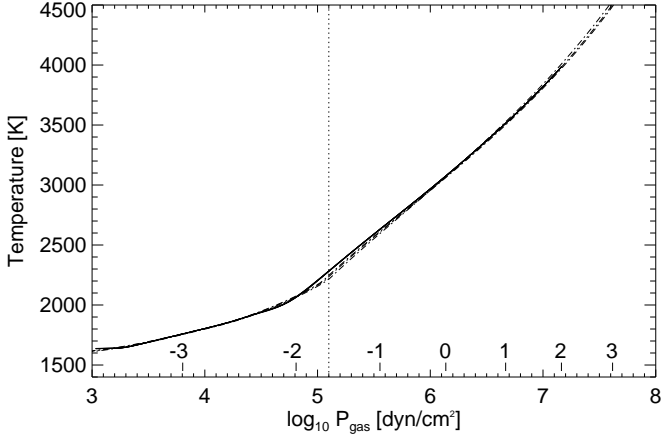}}
\caption[]{%
Overview of the temperature-pressure relation of hydrodynamical
model~\mosm\ (\textbf{solid}) and three standard mixing-length model
atmospheres (\textbf{dash-dotted}) with $\mlp=1.0,1.5,2.0$. The
$\mlp=1.0$ model is the hottest at high pressure while it is coolest
around $\log P=5$. A logarithmic Rosseland optical depth scale is
indicated by the tick marks near the abscissa. All models have
identical atmospheric parameters ($\Teff=2790\pun{K}$, $\logg=5.0$,
solar chemical composition).  The \textbf{dotted} line indicates the
location of the Schwarzschild boundary in the $\mlp=1.0$ model. See
Fig.~\ref{f:talpha2} for an enlargement of the transition region from
convective to radiative transport.
\label{f:talpha1}
} 
\end{figure}

\begin{figure}
\resizebox{\hsize}{!}{\includegraphics[draft = \draftflag]%
{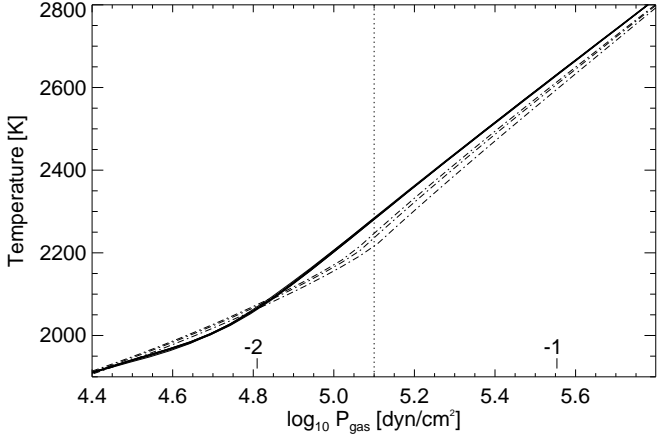}}
\caption[]{%
Temperature-pressure relations like in Fig.~\ref{f:talpha1} but
focusing on the transition region from convective to radiative energy
transport. In this region the $\mlp=1.0$ model is the coolest,
$\mlp=2.0$ the hottest mixing-length model.
\label{f:talpha2}
} 
\end{figure}

\begin{figure}
\resizebox{\hsize}{!}{\includegraphics[draft = \draftflag]%
{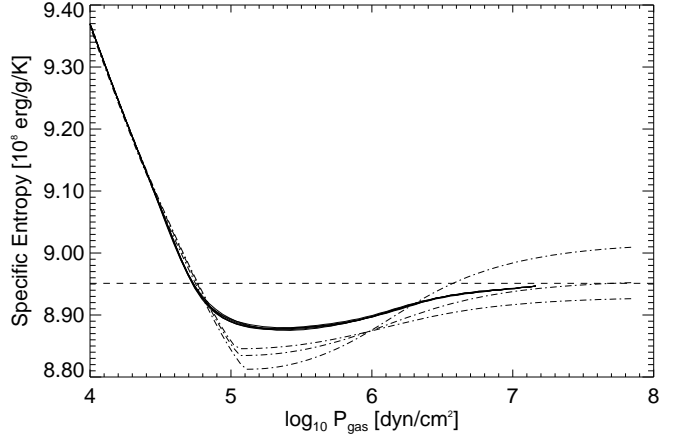}}
\caption[]{%
Entropy-pressure relation of hydrodynamical model~\mosm\
(\textbf{solid}) and three standard mixing-length model atmospheres
(\textbf{dash-dotted}) with $\mlp=1.0,1.5,2.0$. The $\mlp=1.0$ model
has the largest entropy at high pressure while it has the lowest
around $\log P=5$.  The entropy level of material entering the
computational box of hydrodynamical model~\mosm\ at the lower boundary
is indicated by the \textbf{dashed} line.  All models have identical
atmospheric parameters ($\Teff=2790\pun{K}$, $\logg=5.0$, solar
chemical composition).
\label{f:salpha}
} 
\end{figure}

Figures~\ref{f:talpha1}, \ref{f:talpha2}, and \ref{f:salpha} show a
comparison of the thermal structure of our ``best'' hydrodynamical
model~\mosm\ with standard 1D atmosphere models computed with
\LHD. Again, we point out that the treatment of the microphysics
(equation of state and opacities) is equivalent between \MHD\ and
\LHD\ models. Differences lie {\changed mainly} in the dimensionality of the
models --- 3D versus 1D --- and the related MLT treatment of the
convective energy transport in \LHD. 
{\changed In \LHD\ effects of the turbulent pressure due to convective
velocities computed from MLT are neglected. This approximation is well
justified since the turbulent pressure --- as found in the RHD simulations
--- nowhere exceeds 0.5\pun{\%} of the gas pressure.}

For the hydrodynamical model we have plotted in fact six different
mean stratifications in the afore mentioned figures, each one a
horizontal and temporal average over a 250\pun{s} time interval ---
comparable to the turn-over time of the convective cells. The
statistical variations are so small that the profiles are virtually
indistinguishable on the scale of the plots demonstrating again that
we are dealing with atmospheres showing little temporal and horizontal
fluctuations.  Fig.~\ref{f:talpha1} shows that the mean thermal
structure of the hydrodynamical model corresponds closely to those of
mixing-length models.  The mixing-length models themselves do not show
a strong dependence on the mixing-length. In the deeper, convective
layers small horizontal entropy fluctuations
are sufficient to carry the convective flux, and the radiative energy
transport is unimportant.  Here the stratification has to follow
essentially an adiabat in the pressure-temperature plane. The same
reasoning holds for the hydrodynamical as well as MLT models. The
situation is somewhat different in the radiative layers. By
construction, the mixing-length models have to approach a temperature
profile corresponding to radiative equilibrium conditions. Since the
differences among the mixing-length models in the convective region
are small the radiative equilibrium profiles are almost identical as
well. In the hydrodynamical model the temperature of the higher
atmospheric layers ($\log P < 5.3$) is controlled by a competition of
radiative heating and adiabatic cooling of material which overshoots
into regions with formally stable entropy gradient $\frac{ds}{dP}<0$
(see Fig.~\ref{f:salpha}), and, analogously, the radiative cooling and
adiabatic heating of downflowing material. The relative time scales
involved in adiabatic and radiative temperature changes determine how
the temperature will adjust between the upper extreme --- the
radiative equilibrium temperature --- and the lower extreme --- the
temperature achieved during adiabatic expansion of a rising mass
element in the higher atmosphere.

Obviously, in the present case, radiative processes dominate and the
hydrodynamical model closely follows the radiative equilibrium
temperature. As we have seen in the context of the hydrodynamical
models adopting grey \ATLASVI\ opacities this is not necessarily so. In
fact, we trace back the short radiative time scales (see
Fig.~\ref{f:ctimes}) to the presence of molecular absorption which
provides the close coupling of the stratification to the radiative
equilibrium temperature. We speculate here that the situation might
change significantly when one goes to metal poor objects. Convective
velocities are likely not to be reduced as dramatically as the
atmospheric opacities, shifting the relative importance towards
adiabatic cooling. This might lead to a significant deviation from
radiative equilibrium in the atmospheric layers of such
objects. Indeed, for metal poor dwarfs at about solar effective
temperatures such effects are found in hydrodynamical models
\citep[e.g.][]{Asplund+GarciaPerez01}.

Quantitatively, there are small differences in the thermal structure
between the hydrodynamical model and the mixing-length models. As one
can expect the differences are most pronounced in the transition
region between convectively and radiatively dominated energy transport
where the detailed balance between both processes decides about the
resulting temperature profile. Figures~\ref{f:talpha2}
and~\ref{f:salpha} show that the hydrodynamical model has a mean
thermal structure which is close to the profiles of the mixing-length
models but cannot be matched exactly. This certainly comes not as a
big surprise when one considers the simplistic nature of mixing-length
theory. However, certain aspects of the hydrodynamical model can be
matched by choosing a mixing-length model from the set which are parameterized
by \mlp. This is most easily done considering the entropy since it
emphasizes model differences (see Fig.~\ref{f:salpha}). The entropy of
inflowing material of the hydrodynamical model which we interpret as
the entropy of material located in the deep almost perfectly
adiabatically stratified part of the stellar envelope is matched best
by a mixing-length model of $\mlp(\mathrm{evo})=1.5$. In global
stellar structure models the entropy of the envelope controls to some
extent the radius of a star, making $\mlp(\mathrm{evo})$ most relevant
for evolutionary models --- hence, our naming \mbox{``evo''}. For
further justification and discussion of our interpretation see
\citet{Steffen93} and \citet{Ludwig+al99}.

If one wants to match the entropy jump of the hydrodynamical model
$\mlp(\Delta s)=2.1$ provides the closest fit. The entropy jump is the
entropy difference between the atmospheric entropy minimum and the
asymptotic entropy of the deeper layers. {\changed The entropy jump
provides a qualitative measure of the overall available convective
driving}, and in the present case also gives a match to the
temperature gradient of the deeper photosphere with $\log_{10}\tauross
>-1$. The {\em internal\/} uncertainties in the determination of
$\mlp(\mathrm{evo})$ and $\mlp(\Delta s)=2.1$ are small ($\approx\pm
0.05$) as estimated from the statistical fluctuations observed in the
models.
{\changed However, in the present M-dwarf model small deviations from
equilibrium conditions determine the mixing-length parameters.  This
is a challenging situation for any numerical scheme, and we consider
it possible that the actual calibration errors are dominated by
related systematic uncertainties.}

\subsubsection{Importance of the MLT formulation}

Throughout this paper mixing-length parameters are given with respect
to the formulation of MLT according \citet{Mihalas78} (see
\citet{Ludwig+al99} for details of the implementation). The
formulation of Mihalas is widely used, in particular in the context of
stellar atmosphere work. But it is by far not the only MLT ``dialect''
around. Another widely used formulation is the original one by
\citet{BoehmVitense58} which is often encountered in the context of
stellar evolution work. We would like to emphasize that generally the
details in the MLT implementation have a direct and significant
influence on the calibration of the mixing-length parameter. This is
particularly true in the present case where convection penetrates far
into optically thin regions. Typically it is the treatment of the
radiative energy exchange of the convective elements in the optically
thin limit that distinguishes the various MLT formulations.  For
enabling inter-comparison in other context, we like to mention
that the MLT formulation in \PHOENIX\ is similar but slightly different
from the Mihalas formulation. However, test calculations have shown
that numerical differences are negligible in the M-dwarf
regime. Furthermore, in the \ATLASVI\ stellar atmosphere code the
Mihalas formulation is implemented. MLT formulations as given by
\citet{Kippenhahn+Weigert91} and \citet{Stix89} are identical to the
B\"ohm-Vitense formulation.

\begin{figure}
\resizebox{\hsize}{!}{\includegraphics[draft = \draftflag]%
{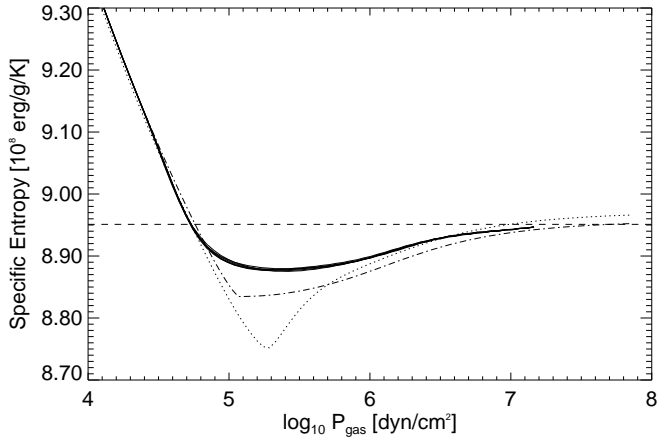}}
\caption[]{%
Entropy-pressure relation of hydrodynamical model~\mosm\
(\textbf{solid}) and standard mixing-length model atmospheres with
$\mlp=1.5$ based on two different formulations of MLT:
\citet{Mihalas78} (\textbf{dot-dashed}) and \citet{BoehmVitense58}
(\textbf{dotted}). Note, the significantly different entropy jumps in
the MLT models employing the {\em same\/} mixing-length parameter. The
entropy level of material entering the computational box of
hydrodynamical model~\mosm\ at the lower boundary is indicated by the
\textbf{dashed} line.  All models have identical atmospheric parameters
($\Teff=2790\pun{K}$, $\logg=5.0$, solar chemical composition).
\label{f:dialect}
} 
\end{figure}

Figure~\ref{f:dialect} shows a comparison of the entropy profiles of
standard mixing-length models of the {\em same\/} $\mlp=1.5$ based on
the the Mihalas and B\"ohm-Vitense formulation. 
{\changed We concentrate on the entropy jump since it is sensitive to
the convective transport in the optically thick and thin layers. The
asymptotic entropy level of the deeper layers is largely controlled by
the entropy change over the optically thick layers only. Mihalas and
B\"ohm-Vitense formulation differ little here, implying the rather
similar outcome. The formulations differ significantly in the
optically thin layers.}  Clearly, the entropy jump comes out to be
almost a factor of~2 larger in the B\"ohm-Vitense formulation! To
reduce the entropy jump to a level comparable as obtained from the
Mihalas formulation, one had to increase the mixing-length parameter
noticeably. Again, this emphasizes that \mlp\ is only well defined
with reference to a specific MLT formulation.

The B\"ohm-Vitense formulation assumes optically thick convective
elements throughout. Their radiative energy exchange is described in
diffusion approximation. It might be surprising that the entropy jump
obtained from the B\"ohm-Vitense formulation is larger than from the
Mihalas formulation which distinguishes between optically thick and
thin elements. The optically thin elements loose (or gain) energy in a
non-diffusive fashion.  A larger entropy jump implies larger radiative
losses of the convective elements. Intuitively, one would expect that
a diffusive transport is rather inefficient, implying a smaller energy
exchange and smaller entropy jump. However, Eq.~\eref{e:tauradii}
shows that the situation is reversed: the thermal adjustment time of
convective elements becomes independent of their optical thickness in
the optically thin limit, while the diffusion approximation predicts
(inaccurately) a decrease of the adjustment time with decreasing
optical thickness. Since the optical thickness becomes very small in
the optically thin limit, the diffusion approximation underestimates
the thermal relaxation time, and {\em overestimates the radiative
losses\/} which leads to a larger entropy jump.

While it appears at first glance arbitrary which MLT formulation one
chooses, Fig.~\ref{f:dialect} also shows that in the present case the
formulation of Mihalas is perhaps superior to the formulation of
B\"ohm-Vitense. The shape of the entropy profile of the Mihalas model
resembles the result of the hydrodynamical model more closely. This
comes at no surprise since the formulation of Mihalas considers
optically thick and thin convective elements --- i.e. better
represents the actual situation. Of course, our choice of relating our
\mlp\ to the formulation of Mihalas was motivated by this property.

\subsubsection{Comparison to the Sun}

{\changed
It is interesting to compare equivalent mixing-length parameters
derived for the thermal structure of the hydrodynamical models with
other ones found in hotter objects, especially the Sun. Using the MLT
formulation of B\"ohm-Vitense, \citet{Ludwig+al99} found in a study of
the convective efficiency based on 2D hydrodynamical models $\mlp=1.6$
for the Sun which has likely to be corrected to $\mlp=1.8$ if one
considers 3D models. In the Sun no distinction needed to be made
between fitting the entropy jump or the asymptotic entropy. The solar
mixing-length models match the photospheric entropy minimum found in
the hydrodynamical models quite well. This is a consequence of the
fact that convection does not reach very far up into optically
thin regions. Together with the relatively large entropy jump which
makes the exact value of the entropy minimum less important this leads
to the ``one $\mlp$ fits both'' situation.  Contrary to the situation
in M-dwarfs differences between the B\"ohm-Vitense's and Mihalas' MLT
formulation are small in the Sun, again, a consequence of the
confinement of convection to optically thick regions. This allows us
to inter-compare the mixing-length parameter obtained by
\citet{Ludwig+al99} for the Sun with the values derived here.
 
The mixing-length parameter of $1.8$ for the Sun and $1.5$ (matching
the asymptotic entropy) or $2.1$ (matching the entropy jump) for the
M-dwarf are not vastly different, the solar value even falls within
the range spanned by the M-dwarf values. In MLT the convective flux
scales as the square of \mlp, hence we are considering a flux
variation within a factor of~2 only.  Convection is efficient in
M-dwarfs in the sense that the whole envelope is almost adiabatically
stratified. The convective energy transport process itself does not
work much differently than in the solar case --- taking the similar
mixing-length parameters as a measure. The adiabatic behavior follows
mostly from the reduction of the transported flux which can be
accommodated with smaller horizontal entropy fluctuations. This
ultimately leads to the smaller entropy jump which amounts to $\Delta
s=7 \times 10^6\pun{erg/g/K}$, or only $\approx 4\pun{\%}$ of the
solar entropy jump. This is remarkable since as evident from
Fig.~\ref{f:salpha} about half of the entropy jump happens to take
place in the optically thin region which is qualitatively different
from the Sun. Moreover, the MLT captures surprisingly well the
convective transport properties under such conditions.}

We note that an extrapolation of the entropy jump as found by
\citet{Ludwig+al99} (their Fig.~4) for F-, G-, and K-dwarfs leads to a
significant overestimation of the entropy jump for M-dwarfs. This
means that the trend in the entropy jump found in hotter objects
changes at some point when moving towards lower \Teff. This, again, is
plausible because the dominant atmospheric opacity source as well as
contributors to the specific heat change in the M-dwarf regime.

\subsection{Synthetic broad band colors}

\begin{table}
\begin{flushleft}
\caption[]{%
Synthetic broad band colors \mbox{V-I}, \mbox{R-I} in the Cousins
system, as well as \mbox{J-K}, and \mbox{J-H} in the CIT system for
the models displayed in Fig.~\ref{f:talpha1}. The effective
temperature~$\Teff^\prime$ corresponds to the total flux as computed
by the detailed spectral synthesis.
\label{t:syncol}}
\begin{tabular}{llllll}
\hline\noalign{\smallskip}
Model           &  $\Teff^\prime$ & \mbox{V-I}& \mbox{R-I} & \mbox{J-K} & \mbox{J-H}\\
\mbox{}         &  [K]            & [mag]     & [mag]      & [mag]      & [mag]\\
\noalign{\smallskip}
\hline\noalign{\smallskip}
\MHD, \mosm       &  2754  & 3.651 & 1.974 & 0.749 & 0.455\\
\LHD, $\mlp=1.0$  &  2743  & 3.733 & 2.010 & 0.739 & 0.449\\
\LHD, $\mlp=1.5$  &  2744  & 3.701 & 1.996 & 0.746 & 0.452\\ 
\LHD, $\mlp=2.0$  &  2745  & 3.686 & 1.989 & 0.749 & 0.454\\
\noalign{\smallskip}
\hline
\end{tabular}%
\end{flushleft}
\end{table}

Table~\ref{t:syncol}  provides  synthetic  optical  and  near-infrared
colors   for    the   models    discussed   before   and    shown   in
Fig.~\ref{f:talpha1}. The colors of the hydrodynamical model have been
calculated  from the average  pressure-temperature structure.   In the
calculation of the colors the  radiation field has been sampled with a
resolution of  2\pun{\AA}, i.e. the treatment  of frequency dependence
is  different  and more  detailed  than the  OBM  used  in the  models
proper. Here  we are  concerned with the  relative behavior  among the
models.   For   the  \mbox{V-I}  and   \mbox{J-K}  colors  appreciable
discrepancies  exist  between observations  and  model predictions  by
\PHOENIX\   (``{\sc  NextGen}''   model  grid)   in   the  \Teff-range
$3600\pun{K}          >          \Teff         >          2300\pun{K}$
\citep{Chabrier+al00,Allard+al01}. In \mbox{V-I} the offset amounts to
$\approx 0.2-0.3\pun{mag}$ in the  sense that the {\sc NextGen} models
are  too blue.  In  \mbox{J-K} the  offset amounts  to $\approx  0.2 -
0.5\pun{mag}$ depending  on the source of water line opacity data 
used in the models.  The  treatment of the convective energy transport
in the  MLT framework ---  employing a standard  \mlp\ of 1.0  --- was
suspected     to     be     one     possible    reason     for     the
discrepancy. Table~\ref{t:syncol}  shows that a  detailed treatment of
convection does  not remedy the situation. The  difference between the
\mbox{V-I} color  of the hydrodynamical model to  the \LHD\ $\mlp=1.0$
amounts to  82\pun{mmag} only.  Moreover, the  hydrodynamical model is
even bluer than  the mixing-length model, i.e. when  assuming the same
differential   behavior  for   models  employing   detailed  radiative
transfer, the  discrepancy between  theory and observations  gets even
slightly  worse.  As  expected from  the thermal  structure $\mlp=2.0$
gives the closest match in color among the three \LHD\ models.

Our results strengthen  those of \citet{Allard+al00} and 
\citet{Chabrier+al00} who
explain most of the visual  discrepancy in terms of missing opacities.
We would also like to emphasize that an optical color mismatch of this
magnitude  does not  indicate  a major  shortcoming  of the  \PHOENIX\
models.  The  V-band, which is likely  to be responsible  for the color
mismatch, is  formed high up  in the atmosphere.  The  radiative energy
flux in the  V-band is a small fraction of  the total radiative output
of the star. Hence, the problem  of the optical colors is related to a
part of  the atmosphere  that is of  minor importance for  the overall
energetics.

Similarly,   the   discrepancies   between  observed   and   predicted
near-infrared colors cannot be  traced back to an inadequate treatment
of  convection   with  MLT.  Possible  differences   between  MLT  and
hydrodynamical models amount to 10\pun{mmag} only.

Table~\ref{t:syncol} allows us to address a side point, namely the
question of how the hydrodynamical model~\mosm\ and the mixing-length
models shown in Fig.~\ref{f:talpha1} can have the same effective
temperature (2790\pun{K}) despite the fact that the hydrodynamical
model is almost always noticebly hotter than the mixing-length
models. Indeed, when calculating the total flux within the detailed
spectral synthesis the corresponding effective temperature
$\Teff^\prime$ of the hydrodynamical model appears to be slightly
($\approx 10\pun{K}$) hotter than the one of the mixing-length models
--- with an offset to the nominal \Teff\ due to the more detailed
treatment of the radiative transfer. We interpret this as an effect of
the small remaining non-linearities when interchanging the averaging
of the temperature-pressure structure with the solution of the
radiative transfer. Strictly speaking, in order to determine the
colors of the hydrodynamical model one must perform the detailed
calculation of the 3D radiation field for many instances in time and
had to average subsequently horizontally and temporally. However, in
the present context, the smallness of the effects did not appear worth
the dramatically increased computational effort.

\subsection{Mean velocity and corresponding effective mixing-length parameter}

\begin{figure}
\resizebox{\hsize}{!}{\includegraphics[draft = \draftflag]%
{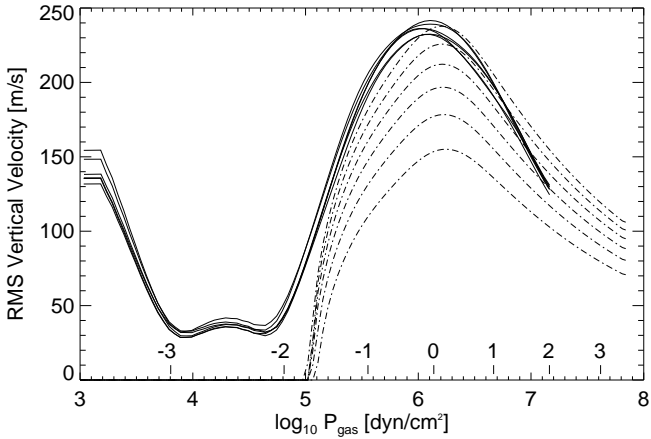}}
\caption[]{%
Velocity-pressure relation of hydrodynamical model~\mosm\
(\textbf{solid}) and five standard mixing-length model atmospheres
(\textbf{dash-dotted}) with $\mlp=1.0$ to $3.5$ in steps of $0.5$.
The velocities of the mixing-length models increase monotonically with
\mlp.  For the hydrodynamical model RMS vertical velocities from six
different temporal windows are plotted. A logarithmic Rosseland
optical depth scale is indicated by the tick marks near the
abscissa. All models have identical atmospheric parameters
($\Teff=2790\pun{K}$, $\logg=5.0$, solar chemical composition).
\label{f:valpha}
} 
\end{figure}

In the previous sections we discussed mixing-length parameters
associated with the thermal properties of our hydrodynamical
models. We now want to turn to a dynamical feature --- the velocity
profile.  Figure~\ref{f:valpha} shows a comparison of velocities of
hydrodynamical model~\mosm\ with convective velocities from
mixing-length models. In the case of the hydrodynamical model RMS
fluctuations of the vertical velocity, for the mixing-length models
the MLT convective velocity are displayed.  Six different temporal
windows (each 250\pun{s} in length) have been used for averaging the
hydrodynamical data, showing noticable variations of the mean velocity
particularly in the higher atmospheric layers. 
{\changed The velocity plateau in the uppermost layers is an
artifact of the upper boundary condition which stipulates the same
velocity at the boundary and the grid points next to the boundary.}
In order to match the
convective velocities in the convectively unstable layers we find that
$\mlp(v)=3.5$ gives a reasonable match around the velocity
maximum. The hydrodynamical model has the tendency to retain higher
velocities towards smaller atmospheric pressure as compared to the MLT
models. The behavior of the velocities towards high pressure shows a
faster decline than the MLT models. The deeper hydrodynamical
model~\mobg\ (not shown) exhibits a similar velocity maximum as
model~\mosm\ but a decline which is not as fast. We conclude that the
decline in model~\mosm\ is to some extend influenced by the lower
boundary condition and an asymptotic behavior towards high pressure
closer to MLT predictions is likely.

The hydrodynamical model shows an appreciable amount of overshoot into
the formally stable atmospheric regions. Of course, by construction,
standard MLT is unable to make any predictions about velocity
amplitudes there. Two different components contribute to the
velocities in the higher atmosphere: one contribution are {\em
waves\/} excited by the stochastical fluid motions in the deeper
layers and travelling upwards, the other contributions are {\em
advective motions\/} overshooting into the stably stratified layers
\citep[see also][]{Ludwig+Nordlund00}.

Qualitatively, we expect that velocity fluctuations $v^\prime$
associated with advective motions decay with increasing distance from
the Schwarzschild boundary, roughly \mbox{$v^\prime \propto P^f$}
where $f$ is a positive parameter of order unity. For undamped sound
waves we would expect a height dependence of roughly \mbox{$v^\prime
\propto P^{-\frac{1}{2}}$}. This means that the relative importance of
both contributions shifts from advection dominated motions to wave
dominated motions with increasing height. If the wave motions are not
strongly damped we would ultimately expect an increase of the velocity
fluctuations with height.  Indeed, according to Fig.~\ref{f:valpha}
this is what we observe in our model. The wave motions mostly arise
from standing waves --- i.e. excited acoustic eigenmodes of the
computational box. The detailed run of the velocity in the overshoot
layers reflects the location of nodes and anti-nodes of the eigenmodes
in vertical direction which explains the non-monotonic behavior of the
atmospheric velocity amplitude.

The spectrum and the structure of the eigenmodes are to some extent
determined by the geometry of the computational domain. This means
that the velocities in the higher atmosphere where the velocity
amplitude is wave dominated are model dependent and should not taken
for real without skepticism. In the next sections we will even argue
that the velocity amplitudes of the acoustic modes observed in our
particular models are probably strongly influenced by the numerics,
and in this sense are mostly artificial.  In order to study the
transport properties of the atmospheric velocity field we will try to
{\em filter out\/} as much of the acoustic contribution to the
velocity field as possible.

\subsection{Atmospheric overshoot and transport time scales}

Stellar atmospheres around $\Teff\approx 2800\pun{K}$ are too hot to
allow for a significant formation of dust grains. However, already at
slightly cooler effective temperatures, grain formation sets in and
dust grains become major opacity contributors, i.e. an important
factor in determining the thermal structure of the atmosphere. In
fact, the spectral energy distribution in the range of effective
temperatures $2500\pun{K} \leq T \leq 1500\pun{K}$ is crucially linked
to the distribution of dust grains in the atmosphere
\citep{Allard+al00}. The amount of dust which is present is determined
by chemical condensation and evaporation processes as sources and
sinks, as well as macroscopic transport processes which carry dust
grains away from their sites of formation. In M-dwarf
atmospheres the transport is dominated by two opposing processes:
gravitational settling of dust grains and their mixing due to the
presence of velocity fields, either related to convection or global
circulations induced by rotation (T.~Guillot 2000, private
communication). In our models, no dust formation takes place but we
nevertheless find it worthwhile to give a characterization of the
atmospheric mixing due to convection and convective overshoot. This
can give at least a first order approximation of how convective mixing
might operate when dust is actually present. The basic question which
we want to address is whether convective overshoot can provide enough
mixing to prevent dust grains from being completely removed from the
atmosphere by gravitational settling.

Formally, we are interested in a statistical representation of the
mean transport properties of the convective velocity field in vertical
direction. The horizontal advection of dust grains by the convective
velocity field probably produces horizontal inhomogeneities in the
dust distribution. We neglect these here since i) we are targeting at
the application of our results in standard 1D stellar atmosphere
models, ii) the horizontal inhomogeneities are small scale (the size
of a convective cell), i.e. hardly observable and iii) the
uncertainties in our understanding of the dust formation process
itself perhaps limits the achievable accuracy anyway. In view of the
last point we present a proxy of the convective mixing only, without
trying to derive a detailed statistical description of the transport
properties of the convective velocity field, i.e.  extracting its
effective transport coefficients \citep[c.f.][ as an example of a more
stringent treatment]{Miesch+al00}.

\begin{figure}
\resizebox{\hsize}{!}{\includegraphics[draft = \draftflag]%
{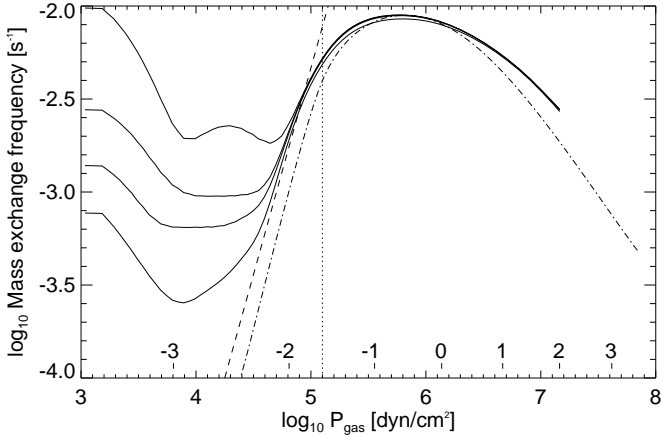}}
\caption[]{%
Mass exchange frequency as a function of pressure for various degrees
of subsonic filtering. The \textbf{solid} curves from top to bottom
correspond to no filtering, $v_\mathrm{phase}\leq 1.0,\, 0.5,\,
0.25\pun{km/s}$, respectively. The \textbf{dashed} line is an
extrapolation towards the asymptotic behaviour corresponding to a
complete removal of acoustic contributions. The \textbf{dashed-dotted}
line depicts the dependence if one assumes a velocity field calculated
from linear theory.  A logarithmic Rosseland optical depth scale is
indicated by the tick marks near the abscissa. The \textbf{dotted}
line indicates the location of the Schwarzschild boundary.
\label{f:massex}
} 
\end{figure}

Figure~\ref{f:massex} shows the mass exchange frequency~\fex\ which
provides an approximate measure of the convective mixing. We define
this quantity as
\beq
\fex(z) \equiv \frac{\langle F_\mathrm{mass}^\mathrm{up}\rangle(z)}{%
                     \langle m_\mathrm{col}\rangle(z)}
\eeq
where $F_\mathrm{mass}^\mathrm{up}$ is the upward directed mass flux,
$m_\mathrm{col}$ the mass column density, and $z$ the geometrical
height. \mbox{$\langle.\rangle$} denotes temporal and horizontal
averages.  Note that the average total mass flux $\langle
F_\mathrm{mass}\rangle$ vanishes
\beq
\langle F_\mathrm{mass}\rangle = \langle\rho u_\mathrm{z}\rangle 
    = \langle F_\mathrm{mass}^\mathrm{up} 
            + F_\mathrm{mass}^\mathrm{down}\rangle = 0.
\eeq
$\rho$ denotes the density, $u_\mathrm{z}$ the vertical velocity
component. \fex\ is the frequency with which the mass above a certain
height~$z$ is replaced by the flux of fresh material from below. Mass
conservation and quasi-steadiness demand that the upward directed mass
flux from below is compensated by a downward directed flux from
above. As long as the flow field is sufficiently disordered this mass
exchange should also go hand in hand with mixing of material; in this
way, \fex\ can be interpreted as mixing frequency of atmospheric layers
above a certain reference height~$z$. The assumption of a disordered
velocity field was checked qualitatively by studying the behaviour of
passive tracer particles being advected with the flow.  The definition
of \fex\ was motivated from our expectation that the amplitude of the
velocity field would strongly decline with distance from the
convectively unstable zone. \citet{Freytag+al96} find in
hydrodynamical models that the velocity amplitude due to convective
overshoot declines exponentially with distance from the Schwarzschild
boundary, and promote this result as a generic feature of overshoot.
In such a situation, the mass flux in or out of a control volume is
dominated by the flux through the surface closest to the convectively
unstable region. This is the reason why we consider just one surface
(or height) through which mass is transported.

As mentioned previously, the atmospheric velocity field has strong
contributions of acoustic waves generated in the deeper layers of the
models. From the comparison (not shown) of the hydrodynamical models
\mosm\ and \mobg\ --- which has a slightly greater extent in depth ---
we concluded that instabilities associated with the lower boundary
condition contribute to the excitation of waves, i.e. the waves are
mostly a numerical artifact. 
{\changed We could not trace down the responsable process, but the behavior was
apparently related to the high degree of adiabaticity the
stratification exhibits close to the lower boundary. In the Sun the
boundary condition operated stably.
Besides the numerical findings,} from a purely theoretical point
of view one would expect that the generation of acoustic energy is
rather inefficient in a flow with as low a Mach number (few percent)
as in the present case. Finally, it turns out that the Richardson
number associated with the horizontal shear introduced by the acoustic
modes in the atmosphere is not low enough to render the shear flow
unstable. No small scale turbulence is induced by the waves and their
oscillatory velocity fields do not contribute significantly to the
mixing of material.  Hence, we concluded that the wave contribution to
the velocity field should be discarded if one wants to measure its
mixing efficiency.

We therefore removed the contribution of waves before evaluating the
average mass flux by subsonic filtering --- a technique invented in
the context of solar observations for cleaning images from ``noise''
stemming from the 5\pun{minute} oscillations \citep{Title+al89}. In
short, one considers a time sequence of images and removes features
with horizontal phase speeds greater than a prescribed threshold. This
is achieved by Fourier filtering of the spatial-temporal data in the
$k$-$\omega$ domain. In our case we worked with 150 snapshots of the
flow field sampled at 10\pun{s} time intervals. The total length of
the sequence was long in comparison to the periods of the excited
acoustic modes. For every depth layer we performed a 3D Fourier
analysis (one temporal, two spatial dimensions) of the vertical mass
flux. We selected velocity thresholds
\beq
v_\mathrm{phase} = \frac{\left|\omega\right|}{\left| \vec{k} \right|} 
     = 1.0,\, 0.5,\, 0.25\pun{km/s}.
\eeq
With these specific choices of $v_\mathrm{phase}$ we tried to remove
acoustic features which had phase speeds in the order of the sound
speed without affecting the slow convective motions which have maximum
velocities of about 0.25\pun{km/s} (see Fig.~\ref{f:valpha}). Note
that the technique accounts only for the horizontal component of the
phase speed. This means that e.g. purely radial oscillations could not
be removed by this procedure. An improvement could, in principle,
extend the filtering technique to four dimensions (one temporal, three
spatial dimensions). But this would require a much higher
computational effort, in particular since one would first have to
determine a suitable set of basis functions for the representation of
features in the vertical direction, where simple harmonic functions do
not form a proper set of eigenfunctions.  Luckily, the standard
subsonic filtering worked well enough in our case so that we did not
have to pursue this issue further.

Figure~\ref{f:massex} shows the resulting mass exchange
frequency~\fex\ as a function of height (expressed in terms of the
mean pressure) for our hydrodynamical model~\mosm. Lowering the
threshold~$v_\mathrm{phase}$ leads to a significant decrease of the
mass exchange frequency in the atmospheric layers while the deeper
layers remain essentially unaffected by the subsonic filtering. This
again shows that the velocity field of the higher atmospheric layers
stems primarily from wave motions. Only the lowest threshold of
0.25\pun{km/s} starts to affect the deeper layers since also
convective flow features reach such velocities. The dashed line in
Fig.~\ref{f:massex} indicates our extrapolation of what to expect in
case of a perfect removal of the acoustic contribution. Its slope is
given by the envelope of the sequence of curves for increasing degree
of subsonic filtering. 
{\changed We want to emphasize that by the very nature of the problem
the filtering procedure does not apply a small correction to the
signal but removes even most of the original signal in some layers.
Wave motions almost inevitably dominate the flow velocity above a
certain atmospheric level. Large corrections as exhibited in
Fig.~\ref{f:massex} are to be expected, but do not pose particular
problems so that we are confident about the filtering procedure.}

As an additional check of our approach we calculated for an LHD
mixing-length model with $\mlp=2.0$ linear eigenmodes describing the
growth of a perturbation under the action of the convective
instability. $\mlp=2$ was chosen since the model gave among the
available mixing-length models the closest match to the thermal
structure of the hydrodynamical model. As horizontal wavelength of the
pertubation we chose the horizontal extent of model~\mosm\ of
250\pun{km}, i.e. the largest horizontal wavelength which could be
accomodated in this hydrodynamical model. \citet{Freytag+al96} showed
that the velocity field of convective eigenmodes gives a reasonable
representation of the velocity field in the overshoot region as long
as the horizontal wavelength is chosen in the vicinity of the actually
present convective scales. The dash-dotted line in Fig.~\ref{f:massex}
shows the shape of \fex\ associated with the linear eigenmode. The
velocity amplitude was scaled to match the maximum of \fex\ in the
deeper layers of the hydrodynamical model.  Besides a systematic shift
the functional behavior predicted by the mode is quite similar to the
one of the hydrodynamical simulation. Moreover, the time scale for the
growth of the perturbation comes out to 120\pun{s} which is of the
same order as the convective turn-over time scale.  The similar rate
of decline of \fex\ in the atmospheric layers strengthens our
confidence in the robustness of our extrapolation procedure.  However,
changing the horizontal wavelength of the perturbation within
reasonable limits as well as experimenting with different background
models did not improve the correspondence between linear modes and
non-linear hydrodynamical model beyond the quality shown in
Fig.~\ref{f:massex}.

Strictly speaking, the mixing frequency displayed in
Fig.~\ref{f:massex} is a lower limit since we removed all
contributions associated with waves. However, the arguments given
before make it very likely that the contributions of wave motions to
the mixing are negligible: the Mach number of the flow is low, and the
weak waves motions which may be excited are associated with little
turbulence.  Therefore we believe that the derived mixing efficiency
is close to what is encountered in a real M-dwarf atmosphere.

What are the possible consequences of the exponential decline of the
mixing time scale due to convective overshoot? Following the work of
\citet{Rossow78}, one arrives at an estimate of about $10^4\pun{s}$
for the typical time scale of dust sedimentation in M-dwarf
atmospheres. According to our hydrodynamical models the convective
mixing could counteract the sedimentation up to a layer of about two
pressure scale heights above the formally convectively unstable
layers. The convective turn-over time scale is of the order of
$10^2\pun{s}$. This is slow enough to allow for dust nucleation in the
upper regions of the convective envelopes of M-dwarfs --- provided
that the thermodynamic conditions allow for nucleation in the first
place. From these considerations we would expect that {\em dust clouds
in M-dwarf (and brown dwarf) atmospheres are confined to layers in the
vicinity of the upper Schwarzschild boundary of the convective
envelope\/}.

\section{Concluding remarks and outlook}

We used elaborate 2D and 3D radiation-hydrodynamics simulations to
study properties of convection on the surface of a prototypical late
M-dwarf ($\Teff\approx 2800\pun{K}$, $\logg=5.0$, solar chemical
composition). Despite the significant differences in the physical
conditions encountered in the solar and an M-dwarf atmosphere we
obtained the striking result that M-dwarf granulation does not look
qualitatively different from what is familiar from the Sun (see
Fig.~\ref{f:flow}). Quantitative differences (intensity contrast
1.1\pun{\%}, horizontal scales $\approx 80\pun{km}$, maximum RMS
velocities $\approx 240\pun{m/s}$, convective turn-over time scale
$\approx 100\pun{s}$) remain within the expectations derived from
mixing-length theory.

Connected to this basic finding is the --- for practical purposes ---
perhaps most important result that the temperature structure of the
higher atmospheric layers is determined by the condition of radiative
equilibrium, and is not very much affected by processes usually not
accounted for in standard stellar atmosphere models. Convective
overshoot as well as energy transport by waves do {\em not\/}
significantly affect the temperature structure outside of the region
of convective instability. We expect that this finding also holds for
main-sequence objects of higher effective temperature where radiation
becomes relatively more important.

Answering the questions posed in the introduction we confirm that in
late M-dwarfs mixing-length theory can be applied to obtain a
realistic description of the convective energy transport in a 1D
stellar atmosphere code, provided one can remove uncertainties related
to the choice of the mixing-length parameter. If an independent
calibration is available we expect that 1D atmosphere models allow
reasonable accurate predictions (on a level as displayed in
Fig.~\ref{f:salpha}) of the atmospheric temperature structure and
ultimately the stellar spectrum to be made. Effects on the stellar
spectrum related to {\em horizontal\/} temperature inhomogeneities are
expected to be very small in the M-dwarf studied here, primarily due
to the small horizontal fluctuations of the thermodynamic variables.
Even in the Sun --- with the much higher temperature contrast present
at its surface --- effects related to horizontal temperature
inhomogeneities are rather subtle \citep[cf.][]{Steffen+Ludwig99}. We
expect that 1D stellar atmosphere models provide an acceptable
overall approximation to the spectrum of main-sequence objects between
the Sun and late M-dwarfs. Only if one demands for a precision
exceeding commonly adopted levels or desires to study effects in
principle not included in standard model atmospheres (e.g. spectral
line shifts and asymmetries), one has to go to more
sophisticated modeling. We emphasize that this refers to objects at
solar metallicity. For metal poor objects the situation is clearly
different \citep[see][]{Asplund+al99b}.

A downside of our findings is that we cannot trace back the remaining
differences between theoretical and observed colors for M-dwarfs to
shortcomings in the treatment of the convective energy transport. The
resolution of the problem has to be found elsewhere, as pointed out
earlier, deficiencies in the molecular opacities are still a possible
option.

We are left with the problem of finding an adequate value of the
mixing-length parameter to obtain a description of the vertical
temperature run in the superadiabatic layers. In the present case we
find a mixing-length parameter \mbox{\mlp($\Delta s$)=2.1} which
gives a match to the entropy jump and the temperature gradient of the
deeper atmospheric layers (see Fig.~\ref{f:salpha}). Despite perhaps
the best value to be employed in stellar atmosphere calculations, for
global stellar structure models a value of \mbox{\mlp($\Delta s$)=1.5}
would be more appropriate since it ensures to find the correct
asymptotic entropy in the convective envelope which is important for
obtaining the correct stellar radius. To complete the ``zoo'' of
mixing-length parameters, we get a value of \mbox{\mlp($v$)=3.5} when
matching the convective velocities predicted in our hydrodynamical
models. We reiterate that all values are given with reference to the
formulation of MLT by \citet{Mihalas78}.

The various values of the mixing-length parameter point towards the
deeper rooted problem that MLT can give a reasonable but not exact
description of the average convective properties. Even calibrating one
aspect does not ensure the overall correct functional form of, say,
the temperature profile. We have seen that different formulations of
MLT can give quite different functional dependencies. They offer the
possibility to improve fits beyond the quality limited by fitting the
mixing-length parameter only.  For late M-dwarfs all this does not
matter much since the differences of the atmospheric structure for
various values of \mlp\ are small. However, we stress that this
statement refers to cooler M-dwarfs on or close to the
main-sequence. For pre-main-sequence (PMS) objects the situation is
markedly different \citep{Baraffe+al02}. There the specific choice of
\mlp\ has a large impact on the resulting atmospheric structure. Work
is underway to extend the present study into this regime which may
also allow us to find the most suitable MLT formulation.

A result beyond the scope of classical model atmospheres is the
derivation of a proxy of atmospheric mixing-time scales (see
Fig~\ref{f:massex}) due to convective overshoot. We find an
exponential ``leaking'' of the convective velocity field into the
formally stably stratified layers. Depending on the exact criterion,
overshooting extends the efficiently mixed regions about 2 pressure
scale heights beyond the Schwarzschild boundary. We suggest that the
mixing found in the $\Teff\approx 2800\pun{K}$ model studied here,
takes place in an analogous fashion in brown dwarfs, and provides the
mixing which counteracts dust sedimentation. Hydrodynamical models can
be used to address this problem more directly by performing
simulations including the formation and transport of dust which we
consider as an interesting challenge for the future.

Last but not least we would like to point out the two weakest points
of our investigation. While the precision of the radiative transfer in
the hydrodynamical calculations is sufficient to address the questions
discussed here there is certainly room for improvement of the OBM to
get an even closer agreement with detailed spectral synthesis
calculations. Secondly, the models presented here are rather
shallow. Improvements in the formulation of the lower boundary
condition would perhaps allow the use of deeper computational domains,
and would reduce the influence of the specific formulation of the
lower boundary conditions.

\begin{acknowledgements}

The authors are indebted to Isabelle Baraffe and Gilles Chabrier for
their supportive enthusiasm during the course of the project, and
their scientific input during many discussions. HGL would like to
thank {\AA}ke Nordlund and Robert Stein for making available a version
of their hydrodynamical atmosphere code. PHH acknowledges support by
``P{\^o}le Scientifique de Mod{\'e}lisation Num{\'e}rique'' (PSMN) at
the {\'E}cole Normale Sup{\'e}rieure de Lyon, NSF grants AST-9720704
and AST-0086246, NASA grants NAG5-8425, NAG5-9222, as well as NASA/JPL
grant 961582 to the University of Georgia (UGA), Athens.  The
calculations presented in this paper were performed on machines
operated by the PSMN , on the NEC-SX5 at the ``Institut du
D{\'e}veloppement et des Ressources en Informatique Scientifique''
(IDRIS), Paris, on the IBM~SP2 of the UGA~UCNS, on the IBM~SP ``Blue
Horizon'' of the San Diego Supercomputer Center (SDSC) with support
from the National Science Foundation, and on the IBM~SP of the NERSC
with support from the US Department of Energy.  We thank all these
institutions for a generous allocation of computer time.

\end{acknowledgements}

\appendix

\section{Computation of the characteristic time scales}

The radiative time scales shown in Fig.~\ref{f:ctimes} were
evaluated in Eddington approximation \citep[see e.g.][]{Edwards90}
according
\beq
t_\mathrm{rad} = \frac{\cp}{16\sigma\chi T^3} \, 
                    \left( 1+\frac{3\chi^2\rho^2}{k^2}\right)
\label{e:tauradi}
\eeq
(\cp\ denotes the specific heat at constant pressure, $\sigma$
Stefan-Boltzmann's constant, $\chi$ opacity, $T$ temperature, $\rho$
mass density, $k$ wavenumber of the disturbance). We assumed that the
geometrical size of a thermal disturbance is the local pressure scale
height~\Hp, hence
\beq
k = \frac{2\pi}{\Hp}.
\eeq
Introducing the optical size of the disturbance
$\tau_\mathrm{dis}\equiv\chi\rho\Hp$ Eq.~\eref{e:tauradi} can be
written as
\beq
t_\mathrm{rad} = \frac{\cp}{16\sigma\chi T^3} \, 
                    \left( 1+\frac{3}{4\pi^2}\,\tau_\mathrm{dis}^2\right).
\label{e:tauradii}
\eeq
Relation~\eref{e:tauradii} shows that in the optically thin limit
$\tau_\mathrm{dis}\ll 1$ the radiative relaxation time is independent
of $\tau_\mathrm{dis}$, and in the optically thick limit
$\tau_\mathrm{dis}\gg 1$ scales quadratically with
$\tau_\mathrm{dis}$.  Equations~\eref{e:tauradi} and \eref{e:tauradii}
describe the decay of a Fourier mode in an infinite isothermal
atmosphere, i.e. make very idealizing assumptions. Our choice of of
the local pressure scale height as size of the perturbation is
somewhat arbitrary. Hence, the reader should take $t_\mathrm{rad}$ as
depicted in Fig.~\ref{f:ctimes} {\changed as order of magnitude
estimates}.

The (adiabatic) Brunt-V\"ais\"al\"a period~$t_\mathrm{BV}$ was
calculated according
\beq
t_\mathrm{BV}=\frac{2\pi}{\sqrt{\left|\omega_\mathrm{BV}^2\right|}},\,\,
\mathrm{with}\,\,
\omega_\mathrm{BV}^2 = \frac{\delta g}{\Hp}\left( \nabla_\mathrm{ad}-\nabla \right).
\eeq 
($\delta\equiv -\left(\pdx{\ln\rho}{\ln T}\right)_\mathrm{P}$ denotes
the thermal expansion coefficient at constant pressure, $g$
gravitational acceleration, $\nabla_\mathrm{ad}$ the adiabatic, and
$\nabla$ the actual temperature gradient of the atmosphere).
The local Kelvin-Helmholtz time scale~$t_\mathrm{KH}$ was evaluated according
\beq
t_\mathrm{KH}  = \frac{P\cp T}{g \sigma\Teff^4} 
\eeq
($P$ denotes the gas pressure) under the assumption that the overall
energy content of a mass column is dominated by its thermal energy
content.

\bibliographystyle{apj}
\bibliography{ms2282}

\end{document}